\documentclass[twocolumn]{autart}    

\usepackage{amsfonts}       
\usepackage{amsmath}
\usepackage{amssymb}
\usepackage{float}  
\usepackage{subfigure}  
\usepackage{url}            
\usepackage{bm}
\usepackage{multirow} 
\usepackage{booktabs}       
\usepackage{algorithmicx}
\usepackage{algpseudocode}
\usepackage{algorithm}
\usepackage{cite}
\usepackage{colortbl,booktabs}
\usepackage{graphicx}          
\allowdisplaybreaks[4]
\newtheorem{lemma}{Lemma}
\newtheorem{definition}{Definition}

\newtheorem{corollary}{Corollary}

\newtheorem{remark}{Remark}
\newtheorem{proposition}{Proposition}
\begin{document}

\begin{frontmatter}

\title{Approximate Dynamic Programming for Constrained Linear Systems: A Piecewise Quadratic Approximation Approach\thanksref{footnoteinfo}} 

\thanks[footnoteinfo]{This paper was not presented at any IFAC 
meeting. Corresponding author Kanghui He.}

\author{Kanghui He}\ead{k.he@tudelft.nl},    
\author{Shengling Shi}\ead{s.shi-3@tudelft.nl},
\author{Ton van den Boom}\ead{a.j.j.vandenBoom@tudelft.nl},               
\author{Bart De Schutter}\ead{b.deschutter@tudelft.nl}  

\address{Delft Center for Systems and Control, Delft University of Technology, Delft, The Netherlands}  


\begin{abstract}                          
Approximate dynamic programming (ADP) faces challenges in dealing with constraints in control problems. Model predictive control (MPC) is, in comparison, well-known for its accommodation of constraints and stability guarantees, although its computation is sometimes prohibitive. This paper introduces an approach combining the two methodologies to overcome their individual limitations. The predictive control law for constrained linear quadratic regulation (CLQR) problems has been proven to be piecewise affine (PWA) while the value function is piecewise quadratic. We exploit these formal results from MPC to design an ADP method for CLQR problems \textcolor{blue}{with a known model}. A novel convex and piecewise quadratic neural network with a local-global architecture is proposed to provide an accurate approximation of the value function, which is used as the cost-to-go function in the online dynamic programming problem. An efficient decomposition algorithm is developed to speed up the online computation. Rigorous stability analysis of the closed-loop system is conducted for the proposed control scheme under the condition that a good approximation of the value function is achieved. Comparative simulations are carried out to demonstrate the potential of the proposed method in terms of online computation and optimality.
\end{abstract}

\end{frontmatter}

\section{Introduction}
\subsection{Background}
Reinforcement learning (RL) \cite{sutton2018reinforcement} provides powerful tools for the synthesis of controllers due to its strong interactions with the environment. When RL is applied to physical systems with given models, model-based RL, also known as (approximate) dynamic programming ((A)DP) \cite{lewis2009reinforcement}, has received much attention since it can make effective use of the state transition information provided by the model. Combined with control techniques, model-based RL has successfully been applied to a broad range of control engineering.

Compared to its diverse industrial applications, the theoretical analysis on the stability and safety for RL faces great challenges \cite{bucsoniu2018reinforcement}. Originated from the artificial intelligence (AI) community, RL used to be developed only for AI applications (e.g., board games), in which AI researchers focused almost entirely on performance with respect to a designated reward function while stability and safety are ignored. Stability and safety, on the other hand, are central to the control community since unstable modes or risky control policies may lead to destructive consequences in physical systems. In the control community, the earliest ADP methods can be traced back to the 1990s, when Werbos \cite{werbos1992approximate} introduced function approximators to construct actor-critic structures for the synthesis of feedback control laws. Thanks to Lyapunov stability theory, the stability issue of ADP has been comprehensively addressed, both for linear \cite{kleinman1968iterative} and nonlinear systems \cite{al2008discrete}. \textcolor{blue}{Recently, \cite{berkenkamp2017safe} has investigated the use of Lyapunov stability results to enforce stability of RL policies during the policy exploration.}

Although ADP approaches consider stability, safety is another important issue that needs further study. From the control viewpoint, safety means that the behavior of closed-loop systems, including states, inputs, and outputs, satisfies some hard or soft constraints. In summary, techniques employed by RL or ADP approaches for dealing with constraints can be grouped into two categories: policy-projection-based methods and policy-optimization-based methods. Policy-projection-based methods consider the RL formulation in the unconstrained case and involves a regulator to check and modify the policies that may violate the constraints. Such regulator can be a single projection operator \cite{gros2020safe,chen2018approximating} that can handle linear state and input constraints, a predictive safety filter \cite{wabersich2021predictive} for nonlinear state and input constraints, or it can contain control barrier functions to guarantee safety \cite{ames2019control,cheng2019end}. These indirect methods, however, sometimes fail to capture the optimal solution of the constrained problem and lack stability guarantees. In comparison, optimization-based methods intend to directly get the optimal value function for the constrained problems by solving the constrained Hamilton-Jacobi-Bellman (HJB) equation. With the optimal value function available, the optimal control policy can thereby be produced by solving a constrained policy optimization problem. \cite{chakrabarty2019approximate} was the first investigation of this kind of method, although it was limited to searching for the best linear feedback control law and needs an initial stabilizing policy. \textcolor{blue}{Following this direction, \cite{chakrabarty2020safe} explores an estimation approach to find an initial stabilizing policy even when there are uncertain nonlinear dynamics.} The authors of \cite{duan2021adaptive} focused on using deep neural networks (NNs) to estimate the value function and the optimal policy for a nonlinear control problem with state constraints.

It should be mentioned that in constrained cases, neither the optimal policy nor the optimal value function is readily available, even for the most basic infinite-horizon linear quadratic regulation (LQR) problems. This to some extent restricts the development of the policy-optimization-based RL methods for constrained control problems. In comparison, model predictive control (MPC) \cite{borrelli2017predictive}, an optimization-based control scheme widely adopted in the control community, has a mature stability and robustness theory as well as an inherent constraint handling. MPC uses multi-step policy optimization without any adaption. From RL's point of view, using multi-step policy optimization algorithms diminishes the need for exact knowledge of the cost-to-go function \cite{bertsekas2019reinforcement}. For infinite-horizon LQR problems, MPC can render the exact optimal control law due to the equivalence of finite and infinite optimal control as long as the horizon is sufficiently long \cite{chmielewski1996constrained}. With this fundamental property, infinite-horizon LQR problems can be solved by either implementing MPC online or explicit MPC \cite{bemporad2002explicit}, and the solution is proven to be piecewise affine (PWA) in the state space \cite{bemporad2002explicit}. However, the computational complexity of online MPC and the storage requirements of explicit MPC grow dramatically with the increase of the horizon. This is one of the main drawbacks of MPC compared with RL or ADP \cite{gorges2017relations}.  

Dedicated to overcoming this drawback, approximation methods of predictive control laws have received much attention in recent years. An emerging methodology is using specific function regression techniques such as polynomial approximators \cite{kvasnica2011stabilizing} and PWA neural networks \cite{chen2018approximating,karg2020efficient,maddalena2019neural}. The PWA neural networks, or more specifically, the neural networks with rectified linear units, can represent PWA functions \cite{montufar2014number}, which provides the opportunity for using PWA neural networks to accurately fit the predictive control laws. In \cite{karg2020efficient,maddalena2019neural}, this approach is shown to alleviate storage demands. Nevertheless, no guarantees of closed-loop stability are conveniently available, even through the final approximation error is small enough. Using NN-based controllers to warm start the solver in online MPC \cite{chen2022large} can inherently guarantee stability, and learning Lyapunov functions to verify the stability of NN-based controllers \cite{chen2020learning} is also an alternative way. However, as additional computation is required in the optimization or learning procedure, the superiority of lower computational complexity brought by approximation methods is not obvious. \textcolor{blue}{\cite{mittal2020neural} presents an on-policy RL framework to learn a Lyapunov NN, which is used in MPC as a terminal cost to generate a safe and stabilizing policy. However, it does not address the online computational issue since the NN can make the MPC problem too complex to solve.} Besides, \textcolor{blue}{\cite{chakrabarty2016support} develops a support vector machine-informed methodology to produce explicit model predictive controllers with feasibility and stability guarantees. To have these guarantees, \cite{chakrabarty2016support} requires sufficient sampling and the initial state to lie in a specific small region.}

\subsection{Novelties and contributions}
\textcolor{blue}{Observing that MPC has computational limitations and approximation in the policy space lacks performance guarantees, we aim to attain a computationally inexpensive control scheme for linear systems with stability and feasibility guarantees. To this end, we approximate the value function via ADP and shorten the prediction horizon to one.} We focus on the infinite-horizon LQR problem with state and input constraints. Note that usually, the optimal control law can be steep in some regions while the value function is continuous and convex \cite{gorges2017relations}. So in general, the value function may be easier to approximate accurately than the control law. Different from the research on approximating the MPC policy \cite{kvasnica2011stabilizing,chen2018approximating,karg2020efficient,maddalena2019neural}, we propose a value function approximation scheme. The optimal value function that has been extensively characterized by explicit MPC, is approximated by a piecewise quadratic (PWQ) neural network. The synthesis of the control law is carried out in a DP problem, where stability, feasibility, and sub-optimality can be guaranteed if a good approximation is obtained. Meanwhile, note that one disadvantage of approximating the value function is that it needs online policy optimization, which would not be the case in policy approximation methods \cite{kvasnica2011stabilizing,chen2018approximating,karg2020efficient,maddalena2019neural}. To address this issue, we develop algorithms so that online optimization can be done efficiently, as will be shown in Section 3. 

The contributions of the paper are highlighted as follows: 

(1) \textcolor{blue}{We propose a novel NN structure to approximate the solution of the constrained LQR problem based on ADP. The proposed NN structure can capture the PWQ and convex properties of the value function. Different from policy-based approximation approaches \cite{chen2018approximating,karg2020efficient,maddalena2020neural}, our approach has the important advantage that the resulting controller has safety and stability guarantees.}

(2) \textcolor{blue}{We propose an efficient algorithm to solve the policy optimization problem, which is a convex piecewise quadratic program. In particular, this program is simplified to a collection of quadratic programs
	(QPs). Note that the main difficulty that prevents ones from considering more complex approximation structures is the increase in online computational time. We solve this problem in Algorithm 1. Complexity analysis and simulation results show that the proposed method requires much less online computation time than implicit MPC.}

(3) \textcolor{blue}{Compared to \cite{chakrabarty2019approximate}, the first exploration of ADP in a constrained LQR setting, the proposed approach eliminates the restriction of searching for a linear feedback law and does not require an initially stabilizing policy nor an initial state belonging to an invariant set.}

(4) \textcolor{blue}{We do a rigorous stability analysis, give stability conditions, and provide a tractable way to verify them. In contrast, \cite{chakrabarty2019approximate} does not address stability, while \cite{bakaravc2018explicit} only explores stability conditions on the bound of the approximation errors but does not provide ways to verify them.}

\section{Preliminaries}
\textbf{Notations:} The identity matrix with dimension $n\times n$ is written as ${{I}_{n}}$. The set $\mathbb{N}$ denotes the set of positive integers containing zero. The symbol $\nabla f\left( \cdot  \right)$ is used to represent the gradient of a function $f\left( \cdot  \right)$. In addition, ${{\lambda }_{\max }}(P )\ \text{and}\ {{\lambda }_{\ \min }}(P)$ represent the maximum and minimum eigenvalues of a symmetric positive definite matrix $P$ respectively. The boundary of the set $ \mathcal{P}$ is $\partial \mathcal{P}$, and $\mathrm{int}(\mathcal{P})$ stands for the interior of $ \mathcal{P}$. We utilize $A_{i,\cdot}$ to represent the $i$th row of the matrix $A$.
\subsection{Infinite-horizon optimal control and MPC}
We study the infinite-horizon constrained linear quadratic regulation (CLQR) problem
\begin{equation}\label{infinite0}
	\begin{aligned}
		J^{*}_\infty(x)=& \min _{U^\infty}  \left\{J_\infty\left(x_0, U^\infty\right) \triangleq \sum_{k=0}^{\infty} x^{T}_{k} Q x_{k} + u^{T}_{k} R u_{k}\right\} \\
		\text {s.t.   }  &x_{k+1}=A x_{k}+B u_{k}, k=0,1,\ldots ,\;x_0 = x\\
		&x_{k} \in \mathcal{X}, u_{k} \in \mathcal{U}, k=0,1,\ldots
	\end{aligned}
\end{equation}
where $x_{k} \in \mathbb{R}^{n}, u_{k} \in \mathbb{R}^{m}, A \in \mathbb{R}^{n \times n}, B \in \mathbb{R}^{n \times m}$, $\mathcal{X}$ and $\mathcal{U}$ are polyhedra that contain the origin in their interior, and $U^\infty = [u_0,u_1,...,u_\infty]$ is the infinite-dimensional decision variable. \textcolor{blue}{Matrices $A$ and $B$ are known. Extending the proposed method to model-free cases will be our future work. In addition, computing $K^*$ and $P^*$ is also possible when the model is unknown \cite{chakrabarty2019approximate}.} Like other papers \cite{kerrigan2000invariant,chen2018approximating}, it is assumed that 

\textbf{Assumption A1:} $(A, B)$ is stabilizable, $Q >0$, and $R>0$. Moreover, there exists an initial state, such that there exists a sequence of admissible input vectors $u_0,\;u_1,\dots$ that can steer the state to the origin, i.e., $\bar{X} \triangleq \left\{ x \in {\mathbb{R}^n}|\exists  U^\infty \;{\text{s}}{\text{.t}}{\text{.}}\;x_k \in \mathcal{X},\;u_k \in \mathcal{U},\;{\text{and}}\;J^{*}_\infty(x)\right.$ $\left.{< \infty ,\;\forall k \in \mathbb{N}\;} \right\}$ is not empty.

In some existing literature \cite{kerrigan2000invariant,borrelli2017predictive}, $\bar{X}$ is also called the maximal stabilizable set.

With Assumption A1, let $K \in \mathbb{R}^{m \times n} $ be a stabilizing gain matrix for the unconstrained plant $x_{k+1}=A x_{k}+B u_{k}$. As in \cite{chmielewski1996constrained}, we consider the admissible set of states under a given stabilizing control gain $K$ as $\bar{\mathcal{X}}_{K}=\left\{x \in \mathbb{R}^{n} \mid x \in \mathcal{X},-K x \in \mathcal{U}\right\}$.

If the constraints in the CLQR problem \eqref{infinite0} are removed, the problem admits a unique linear solution $u^{\text{LQR}}_k=-K^*x_k, \;k=0,1,\ldots$ where $K^{*}=\left(R+B^{T} P^{*} B\right)^{-1} B^{T} P^{*} A$ and $P^* = P^{*T}$ is the unique positive-definite solution of
the algebraic Riccati equation \cite{kleinman1968iterative}
\begin{equation}\label{are}
	P^{*}\!=\!A^{T} P^{*} A\!-\!A^{T} P^{*} B\left(R\!+\!B^{T} P B\right)^{-1} B^{T} P^{*} A+Q
\end{equation}
\textcolor{blue}{Then, $J_{\infty}^*(x) = x^TP^*x$ in the absence of constraints.}

In the constrained case, problem \eqref{infinite0} is not as easy to solve as the unconstrained one due to the infinite number of decision variables. However, if the state is close to the origin, the unconstrained LQR solution $u^{\text{LQR}}_k=-K^*x_k, \;k=0,1,\ldots$ may not violate the constraints so that the solution to problem \eqref{infinite0} is identical to the unconstrained LQR solution $-K^{*} x_{k}$ as if there is no constraint at all. This motivates researchers to consider the following positively invariant set:
\begin{definition}[Maximal LQR invariant set \cite{borrelli2017predictive,chmielewski1996constrained}]
	 $\mathcal{O}^\text{LQR}_{\infty} =\{{x \in \mathbb{R}^{n} \mid\left(A-BK^*\right)^{k} x}$ $\left.{\in \bar{\mathcal{X}}_{K^*}, \forall k \in  \mathbb{N}  \subseteq \mathbb{R}^{n}}\right\}$ is called the maximal LQR invariant set for the autonomous constrained linear system $x_{k+1}=\left(A-BK^* \right) x_k$, $x_k \in \mathcal{X}, \forall k \in  \mathbb{N}$.
\end{definition}
With this definition, the existing literature \cite{chmielewski1996constrained,scokaert1998constrained,pang2020data} considers using a finite-horizon problem:
\begin{align}\label{e5}
	J_{N}^{*}(x)=&\min _{\left\{u_{k}\right\}_{k=0}^{N-1}}  \sum_{k=0}^{N-1} x^T_k Q x_k + u^T_k R u_k +x^T_N P^* x_N\nonumber\\
	\text { s.t.} \quad & x_{k+1}=Ax_k+Bu_k, \; k=0,1, \ldots,N-1, x_0 = x\nonumber\\
	& x_k \in \mathcal{X},\; u_k \in \mathcal{U},\; k=0,1, \ldots ,N-1
\end{align}
to approximate the infinite-horizon LQR problem (1). In \eqref{e5}, $U = [u_0,u_1,...,u_{N-1}]$ is the decision variable and $U^{*} (x)$ is the solution. The rationale behind this design is that after a sufficient long horizon, the resulting $x_N$ will fall into $\mathcal{O}^\text{LQR}_{\infty}$ where the unconstrained LQR solution $u_N = -K^* x_N$ will not violate the constraints. Since $\mathcal{O}^\text{LQR}_{\infty}$ is positively invariant, the subsequent trajectories $x_{N+1}, x_{N+2},\dots$ will always stay in $\mathcal{O}^\text{LQR}_{\infty}$, and the constraints will be automatically satisfied with the control gain $K^*$. As the result, the terminal cost $x^T_N P^* x_N$ is an exact representation for $J_{\infty}^{*}\left(x_{N}\right)$. The following theorem, proposed in \cite{chmielewski1996constrained}, formally demonstrates the validity of this design.

\begin{thm}[\cite{chmielewski1996constrained}]\label{t1}

 Let $\bar{J}$ be an upper bound on $J^*_\infty(x)$. For any $x \in \bar{X}$, if $N > \left( {\bar J - p} \right)/q$ where $0<q \leq q_\text{m} \triangleq \inf _{x \notin{\mathcal{O}^\text{LQR}_{\infty}}}\left\{x^{\mathrm{T}} Q x\right\}$ and $0<p \leq p_\text{m} \hat{=} \inf _{x \notin \mathcal{O}^\text{LQR}_{\infty}}\left\{x^{\mathrm{T}} P x\right\}$, then $x_N \in \mathcal{O}^\text{LQR}_{\infty}$ and problem \eqref{e5} is equivalent to problem (1) in the sense of $J_{\infty}^*(x) = J_N^*(x),\; \forall x \in \bar{X}$.
\end{thm}

According to Theorem \ref{t1}, the horizon $N$ is essential to the equivalence of the finite-horizon and infinite-horizon problems. For a given $x$, one can repeatedly solve problem \eqref{e5} with an increasing $N$ until the terminal state falls into $\mathcal{O}^\text{LQR}_{\infty}$ \cite{scokaert1998constrained}. This approach only works for a certain state value. Alternatively, \cite{chmielewski1996constrained} gives a conservative estimate of the horizon for all $x$ belonging to a polyhedral set $X_0$ based on the result in Theorem \ref{t1}. Specifically, \cite{chmielewski1996constrained} computes $p_\text{m}$, $q_\text{m}$, and $J_{\infty}^{*}(\cdot)$ at the vertices of $X_0$. Next, $\bar{J}$ is set to be maximum of all $J_{\infty}^{*}(\cdot)$, and $N$ is selected to be the smallest integer such that $N\geq(\bar{J}-p_\text{m})/q_\text{m}$.

It is not necessary to compute $\mathcal{O}^\text{LQR}_{\infty}$ explicitly. A more mathematically convenient routine is to compute an ellipsoidal subset of $\mathcal{O}^\text{LQR}_{\infty}$ \cite{chmielewski1996constrained}. This approach, on the other hand, may lead to a conservative choice of $N$.

With the equivalence in Theorem \ref{t1} established, we can regulate the state to minimize the infinite-horizon cost by solving problem \eqref{e5}. Throughout the paper, we assume that $N$ in \eqref{e5} is chosen such that the equivalence in Theorem \ref{t1} is satisfied. Problem \eqref{e5} can be solved in a receding-horizon manner. Specifically, at each time step $t, t \in \mathbb{N}$, $x$ in \eqref{e5} is set to the current state $x_t$ and only the first $m$ elements of the corresponding solution $U^{*} (x_t)$ is applied to the system, i.e., $u_t^*({x_t}) = [{I_m}\;\;\;\underbrace {{0_m}\;\;\; \cdots \;\;\;{0_m}}_{N - 1\;{\text{terms}}}]U^{*} (x_t)$. Then, problem \eqref{e5} is a typical MPC problem.

By substituting the state update equation into $J_{N}^{*}(x)$ and the constraints, \eqref{e5} can be reformulated as a multi-parametric quadratic program (mpQP) that only depends on the current state $x$:
\begin{equation}\label{empc}
	\begin{aligned}
		J^*_N(x)=&\frac{1}{2} x^T_0 Y x_0+\min _{U}\left\{\frac{1}{2} U^{T} H U+x^T_0 F U\right\}.\\
		&\text { s.t. } \quad G U \leq w+Sx_0, x_0=x;
	\end{aligned}
\end{equation}
where the matrices $H, F, Y, G, w, S$ can be readily obtained through matrix multiplication operations. Problem \eqref{empc} can be solved online or offline. The online implementation requires to solve a quadratic program (QP) with $x$ in \eqref{empc} fixed to the current state at each time step. Although efficient QP solvers based on interior point methods and active-set methods are applicable, the computational complexity is cubic in the horizon $N$ and the dimension $m$ of the input vector. In comparison, the offline method, also called explicit MPC, aims to directly solve the mpQP in \eqref{empc} for all possible $x$ in a feasible set, and find the relationship between the optimizer $U^*(x)$ and $x$. 

The results from explicit MPC show that the optimizer $U^*(x)$ and the optimal MPC control law are PWA. Therefore, the solution to the infinite-horizon problem \eqref{infinite0} is also PWA. The following theorem, proposed in \cite{bemporad2002explicit}, summarizes the properties of the explicit solution to the CLQR problem. 

\begin{thm}[\cite{bemporad2002explicit}]\label{pwaforlqr}
	 With Assumption A1 and the equivalence in Theorem \ref{t1} satisfied, the state feedback solution to the CLQR problem \eqref{infinite0} in a compact polyhedral set of the initial conditions $X_0  \subseteq \bar{X}$ is time-invariant, continuous and PWA
	\begin{equation}\label{pwalqr}
		u^{*}=F_{j} x+g_{j} \quad \text{if} \quad x \in \mathcal{R}_{j}, \quad j=1, \ldots, N_\mathrm{r}
	\end{equation}
	where the polyhedral sets $\mathcal{R}_{j}=\left\{x \in \mathbb{R}^{n} : \; H_{j} x \leq h_{j}\right\}, j=1,\dots,N_\mathrm{r}$ constitute a finite partition of $X_0$.
\end{thm}

To implement the PWA feedback law, the direct way is to store $H_j,h_j,F_j,g_j$ that define all the polyhedra $\mathcal{R}_{j}$ and the corresponding affine feedback law, perform an online search to locate the polyhedron that contains $x$, and finally feed the system with the corresponding affine feedback law.

\subsection{Problem formulation}
The explicit controller is easy to compute offline in the case of a short horizon and a low-dimensional input vector. However, with the increase of the horizon and the input's dimension, the number of regions in \eqref{pwalqr} grows exponentially \cite{bemporad2002explicit} and the representations of these regions become more complex, which may make the offline computation and online implementation intractable. 

It has been verified that artificial neural networks with at least one hidden layer have the capability of universal approximation \cite{barron1993universal} for any continuous function, provided that the hidden layer is given enough units. Herein, one direct intuition is that we can use neural networks to represent the explicit MPC law, without any need to identify the regions in \eqref{pwalqr}. With the knowledge of the PWA form of the MPC law, one can construct some specific neural architectures to achieve exact and computationally inexpensive approximation. Related work has been reported in \cite{karg2020efficient,chen2018approximating,maddalena2020neural}, combined with supervised learning or policy gradient methods. The neural-network-based controllers proposed in these papers require online feasibility certificate to determine whether the outputs of the neural networks are safe. Usually, a feasibility recovery technique based on projection is used to regulate the control inputs. These neural network based policy approximators, on the other hand, inherently lack stability guarantees. The authors of \cite{chen2022large} propose an explicit-implicit MPC scheme, where neural networks provide a control input initialization and a primal active-set method is executed to solve a QP online to ensure feasibility and asymptotic stability. However, this explicit-implicit MPC scheme needs large-scale neural networks whose outputs are supposed to contain all primal and dual variables in \eqref{empc}, and the online computation burden still exists since the decision variable in the QP is still $U$. 

\textcolor{blue}{To address the computational burden of MPC and the lack of guarantees of policy-based approximation, we adopt value function approximation and shorten the MPC horizon to one. The main challenges are thereby (i) the design of the value function approximator, which is expected to output a function akin to the optimal value function, (ii) the relation between those guarantees and the quality of the approximation, and (iii) further reduction of online computation time concerning that the one-step problem contains a function approximator.}

Before establishing our approximation structure, we concentrate on the properties of the optimal value function $J_{\infty}^{*}(\cdot)$.

\begin{thm}[\cite{borrelli2017predictive}]\label{t2}
	 With Assumption A1 and the equivalence in Theorem \ref{t1} satisfied, then, in a compact polyhedral set of the initial conditions $X_0  \subseteq \bar{X}$, the optimal value function $J_{\infty}^{*}(\cdot)$ obtained from $J_{N}^{*}(\cdot)$ is continuous, convex, and PWQ over polyhedra:
	\begin{equation}\label{e9}
		J^{*}_\infty(x)=J_i(x)=x^{T} P_i x+q^T_{i} x+v_i, \text { if } x \in \mathcal{R}_i, i=1, \ldots, N_\mathrm{r}
	\end{equation}
	Moreover, if the mpQP problem \eqref{empc} is not degenerate, then the value function $J_{\infty}^{*}(\cdot)$ is continuously differentiable.
\end{thm}

\begin{thm}[\cite{baotic2008efficient}]\label{t3}
	 Assume that the infinite-horizon problem \eqref{infinite0} results in a non-degenerate mpQP. Let $\mathcal{R}_i,\mathcal{R}_j$ be two neighboring polyhedra and $\mathcal{A}_i,\mathcal{A}_j$ be the corresponding sets of active constraints at the optimum of \eqref{empc}, i.e., $\mathcal{A}_i=\left\{l \mid G_{l,\cdot} U^{*}(x)=w_{l, \cdot}+S_{l, \cdot} x, x \in \mathcal{R}_i \right\}$. Then,
	\begin{align}\label{P}
		&P_i-P_j\leq 0 \text{ if } \mathcal{A}_i \subset \mathcal{A}_j\nonumber\\
		&P^{*}-P_{j} \leq 0 \quad \forall j \in \left\{{1,2,\dots,N_\mathrm{r}}\right\}
	\end{align}
\end{thm}
Problem \eqref{empc} is said to be (primal) degenerate if $G_{\mathcal{A}_{j},\cdot}$ is not full row rank in some regions $\mathcal{R}_{j}$. Note that some CLQR problems may result in degenerate mpQPs, which means the continuous differentiability of $J_{\infty}^{*}(\cdot)$ may not always hold. For example, degeneracy could happen when more than $Nm$ constraints in \eqref{empc} are active at the optimizer $U^{*}(x)$.  Nevertheless, degeneracy does not necessarily cause the loss of continuous differentiability of the value function \cite{baotic2008efficient}. Also notice that currently, there are no analytical results on how to identify a class of control problems that can ensure non-degeneracy in advance.  Adding a terminal constraint may make \eqref{empc} become degenerate \cite{baotic2008efficient}. If degeneracy is recorded when solving \eqref{empc}, feasible methods to avoid degeneracy include removing redundant constraints from $G U \leq w+S x_{0}$ \cite{tondel2003algorithm} or slightly tuning the weight matrices $Q$ and $R$.

\section{ADP design for CLQR problem}
\subsection{Design of neural network for approximation in value space}
According to Theorems \ref{t2} and \ref{t3}, the value function approximator, denoted by $\hat{J}(\cdot, \theta)$ where $\theta$ refers to some parameters, is expected to have the following features: 
\begin{itemize}
	\item (F1) It can partition its input space into polyhedral regions.
	\item (F2) It can produce a convex and PWQ function partitioned by polyhedra.
	\item (F3) For $x$ in a small region containing the origin, the approximator can provide the exact representation for the value function, i.e., $\hat{J}(x, \theta) = x^TP^*x$.
\end{itemize}
We intend to use a feed-forward NN to capture the relationship between the state $x$ and the value $J_\infty^*(\cdot)$ of the optimal value function. A feed-forward NN is composed of one or more hidden layers and an output layer, where each hidden layer contains an affine map
$ f_{l}\left(\kappa_{l-1}\right)=W_{l} \kappa_{l-1}+b_{l}$, followed by a nonlinear map $\kappa_l = g_l(f_l)$. Here, $M_l$ is the width of the $l$th layer, referring to the number of units in the layer, $M_0=n$, $\kappa_{l-1} \in \mathbb{R}^{M_l}$ is the output of the previous layer with $\kappa_{0}$ the input of the network, $W_l \in \mathbb{R}^{M_l\times M_{l-1}}$ and $b_l \in \mathbb{R}^{M_l}$ are the weights and biases, respectively, $g_l(\cdot) : \mathbb{R}^{M_l} \to \mathbb{R}^{M_l}$ is a nonlinear activation function that is usually applied element-wise on its input.

In the output layer, the outputs of the last hidden layer are linearly combined with weight $W_{L+1} \in \mathbb{R}^{1 \times M_l}$ and bias $b_{L+1} \in \mathbb{R} $, to produce the final output of the network. Based on these definitions, a NN with $L$ hidden layers and $M_l$ units in the $l$th layer can be represented by 
\begin{equation}\label{nn}
	f_\mathrm{NN}(x,\theta)=[f_{L+1} \circ g_{L} \circ f_{L} \circ \cdots \circ g_{1} \circ f_{1}](x)
\end{equation}
where $\theta$ contains all the weights and biases of the affine functions in all the layers, and the symbol $\circ$ means the layers are connected in series.

The activation function plays an important role in the approximation of NNs. In this paper, we consider a popular activation function, named the rectifier linear unit (ReLU), which is defined as $g_\text{ReLU}(x) = \max \left\{ {0,\;x } \right\}$. Here, $\max(0,\;x)$ computes the element-wise maximum between the vector $x$ and a zero vector of the same dimension. An amazing property of the ReLU is that it can produce a series of PWA functions with polyhedral partitions, combined with an affine transformation \cite{montufar2014number}. Actually, the output of a NN with ReLUs as activation functions is PWA. Based on these observations, \cite{karg2020efficient,chen2018approximating,maddalena2020neural} contemplate using ReLU NNs to represent the explicit MPC law. 

As the optimal value function is PWQ, we are interested in producing a class of PWQ basis functions that can efficiently represent the value function. Let us focus on the last hidden layer in \eqref{nn}. All activation units in this layer can still be written as a continuous PWA function over the input space of the network \cite{fahandezh2020proximity}. \textcolor{blue}{The element-wise product of any two vector-valued continuous PWA functions with the same number of components is a continuous PWQ function.} This indicates that such a product can be a possible basis to capture the features of the MPC optimal value function. We therefore calculate the element-wise product $\odot$ of all the units $\kappa_L$ in the last hidden layer
\begin{equation*}\label{product}
	\textcolor{blue}{\varphi = p\left(\kappa_{L}\right) \triangleq \text{diag}(\kappa_{L})  \kappa_{L} \in \mathbb{R}^{M_L}}
\end{equation*}
to generate a series of PWQ functions $\varphi=[\varphi_1(x)\;\;\varphi_2(x)$ $\cdots\;\;\varphi_M(x)]^T$. This calculation can be viewed as a layer in the network, denoted by the product layer $p(\kappa_L)$. 

Finally, the output layer is a simple weighted sum of the outputs of the previous layer:
\begin{equation*}\label{output}
	f_{L+1}(\varphi) = r^T\varphi  = {\sum\limits_{i = 1}^{M_L} {{r_{i}}{\varphi _{i}}} } 
\end{equation*}
with the weight vector $r = {\left( {{r_{1}},\;{r_{2}},\;...,\;{r_{M_L}}} \right)^T} \in {\mathbb{R}^{M_L}}$. 

To make the proposed approximator satisfy F3, we develop a “local-global” architecture \cite{bertsekas1996neuro}, which decomposes the outputs of the approximator into two parts
\begin{equation}\label{nn2}
	\hat J\left( {x,W,b,r} \right) = x^T P^*x + [f_{2} \circ p \circ g_{1} \circ f_{1} ](x)
\end{equation}
with $P^*$ the solution to the algebraic Riccati equation. The term $x^T P^*x$ is included to capture “global” aspects of $J_{\infty}^{*}(\cdot)$, while the neural-network-based term is exploited to identify the polyhedral partition in \eqref{e9} and capture the local residuals $J_{\infty}^{*}(x)-x^T P^*x$. Since the known term $x^T P^*x$ that dominates the value function is extracted and fixed, using such a “local-global" architecture is possible to enhance the quality of approximation.

We hereafter denote $M_1$ by $M$, and $	\hat J\left( {\cdot,W,b,r} \right)$ by $\hat J\left( {\cdot,\theta} \right)$ for the sake of brevity, with all the parameters condensed in $\theta$.

The reason why we use only one hidden layer is twofold. First, by using one hidden layer, it is possible to construct a convex $\hat J\left( {\cdot,\theta} \right)$, which is desirable since the real optimal value function is convex. This property will be discussed in the next section. More importantly,  using one hidden layer provides the opportunity to solve the dynamic programming problem through simple QP methods. It should be noticed that in our case the dynamic programming problem is a nonlinear program since the objective function is PWQ. If we use one hidden layer, we can explicitly compute the coefficients of $\hat J\left( {\cdot,\theta} \right)$ by extracting the activated units in the hidden layer.  The online dynamic programming can be conducted by solving a sequence of quadratic programs without the need to calculate any gradients or Hessians.

\subsection{Network training and convexity analysis}
So far, we have finished the design of the architecture of the proposed neural network. This subsection discusses the optimization strategy over the parameters in $\theta$ in order to approximate the optimal value function. 

Firstly, training data needs to be generated offline by solving  \eqref{e5} for $N_x$ different initial states $\left\{ {{x^{(i)}}} \right\}_{i = 1}^{{N_x}}$, $x^{(i)} \in X_0$, and get the state-value pairs $\{(x^{(i)}, J_{\infty}^{*}(x^{(i)}))\}^{N_x}_{i=1}$.  Let ${U^{{(i)}*}} =\{ {u_k^{(i)*}} \}_{k = 0}^{N - 1}\;{\text{and}}\;\{ {x_k^{(i)*}} \}_{k = 0}^N$ denote the solution to the MPC problem for the initial state ${x^{(i)}}$ and the corresponding trajectories of the closed-loop controlled system. All existing research \cite{karg2020efficient,maddalena2020neural} on the approximation of MPC only uses the initial states $\{ {x_0^{(i)*}} \}_{i = 1}^{{N_x}}$ as the input training data, which means solving \eqref{e5} ${{N_x}}$ times can only generate ${{N_x}}$ training pairs. 

In view of this, we present a more efficient strategy for data generation by leveraging the equivalence between the finite-horizon problem and the infinite-horizon problem. Suppose that we have obtained the optimal control sequence $\{ {u_k^{(i)*}} \}_{k = 0}^{N - 1}$ and the corresponding value $J_{N}^{*}\left(x^{(i) }\right)$ for different $x^{(i) }$, consider the sub-problems whereby we start at ${x_k^{(i)*}},\; \text{ for } \; k =1,\dots,N-1$ and wish to minimize $J_{\infty}(x_k^{(i)*}, U)$ in \eqref{infinite0}. According to the principle of optimality \cite{bertsekas2019reinforcement}, the truncated optimal control sequence $\{ {u_j^{(i)*}} \}_{j = k}^{N-1 }$ is also optimal for these sub-problems. As a result, the optimal value functions for these subsequent trajectories ${x_k^{(i)*}},\; k =1,\dots,N-1$ can directly be computed as
\begin{align}\label{sufficient}
	J_\infty ^*( {x_k^{(i)*}} ) &= J_N^*( {x_k^{(i)*}}) \nonumber\\
	&= J_N^*( {x_{}^{(i)}} ) - \sum\limits_{j = 0}^{k - 1} {x_j^{(i)*T}Qx_j^{(i)*} + u_j^{(i)*T}Ru_j^{(i)*}} 
\end{align}
with no need to solve \eqref{e5} repeatedly. 

With this design, we can generate $N_xN$ state-value pairs by only solving \eqref{e5} $N_x$ times. \textcolor{blue}{Therefore, our sampling strategy is sampling from the MPC closed-loop trajectories. Besides, since we have assumed the complete knowledge of the model, data (state-value pairs) can be rapidly and accurately generated once the initial states $x_0$ are sampled. } In addition, for any $x^{(i)}$, the control sequence $\{u_{k}^{(i) *}\}_{k=0}^{N-1}$ will regulate the state to a point in $\mathcal{O}_{\infty}^\text{LQR}$ and the subsequent states will always stay in $\mathcal{O}_{\infty}^\text{LQR}$ according to Theorem \ref{t1}. This means that the tail of the state sequence $\{x_{k}^{(i) *}\}_{k=0}^{N-1}$ is prone to congregating near the origin if the initial state $x^{(i)}$ is chosen too close to the origin. Therefore, to make the training data cover the whole region of $X_0$ as much as possible, the initial states $x^{(i)}, i=1,\dots,N_x$ are recommended to be set near the boundary of $X_0$.

With the $N_xN$ state-value pairs available, the neural network is trained so that its parameters approximate the solution to the following problem:
\begin{equation}\label{training}
	\begin{gathered}
		\mathop {\min }\limits_{b<0,W,r\geq 0} \frac{1}{{N{N_x}}}\sum\limits_{i = 1}^{{N_x}} {\sum\limits_{k = 0}^{N - 1} {e( {{x^{ (i)*}_k},\theta } )}} \;\;\; \hfill \\
	\end{gathered} 
\end{equation}
where $e( {{x^{ (i)*}_k},\theta } ) = (\hat{J}(x_{k}^{(i) *}, \theta)-J_{\infty}^{*}(x_{k}^{(i) *}))^{2}$ is the square of the approximation error for each training pair, and $x_0^{(i)*} = x^{(i)},\; \forall i\in \left\{1,…,N_x\right\}$. The constraint $b<0$ is introduced to guarantee that no units in the hidden layer are activated when $x$ is near the origin, i.e., to fulfill F3, while the constraint $r \geq 0$ is responsible for maintaining convexity of $\hat{J}(\cdot, \theta)$.

Problem \eqref{training} is a nonlinear least-squares problem, which can be successfully solved by the gradient descent method \cite{goodfellow2016deep} with a learning rate $\alpha >0$. \textcolor{blue}{The constraints on the NN parameters can be handled by constraint elimination, i.e., by letting $r=\left(\bar{r}^2_1, \bar{r}^2_2, \ldots, \bar{r}^2_M\right)^T$, which can always guarantee $r \geq 0$, penalizing the constraint violation in the loss function, or reducing the number of hidden units if constraint violation is detected.}


After the neural network is trained, the system can be run and the control signals are computed by solving a dynamic programming problem. It is desirable that it could be convex so that any locally optimal point is also globally optimal. It should be noticed that in most case a neural network cannot provide a convex function w.r.t. its inputs even through it uses convex activation functions. With a specific structure, our proposed neural network, on the other hand, allows to produce a convex function $\hat{J}(\cdot, \theta)$.

Suppose that the output of the proposed approximator has the following PWQ form:
\begin{equation}\label{e26}
	\hat{J}(x, \theta) \!=\!\hat{J}^j(x)\! =\!x^{T} \hat{P}_j x+\hat{q}^{T}_j x+\hat{v}_j, \text {if} \;x \in \mathcal{\hat{R}}_j, j\!=\!1, \ldots, \hat{N}_{\mathrm{r}}
\end{equation}
where $\hat{\mathcal{R}}_{j}$ are polyhedra defined by the hyperplanes $\{W_{l, \cdot}x+b_{l}=0\}^M_{l=1}$. Define $\bar{R}=\text{diag}(r)$ and we can rewrite \eqref{nn2} as 
\begin{equation}\label{e27}
	\hat{J}(x, \theta)=x^T P^*x+ \kappa^T \bar{R} \kappa
\end{equation}
where $\kappa$, the output of the  hidden layer, is PWA w.r.t. $x$. Therefore, by applying the chain rule for the second derivative, it is found that the Hessian $\nabla_x^2\hat J$ is positive definite in the interior of any $\hat{\mathcal{R}}_j$ if the weight matrix $\bar{R} \geq 0 $, i.e., $r_i \geq 0, \text{ for } i=1,\dots,M$. In the following, we will show that the gradient descent method with projection can guarantee the positive semi-definiteness of $\bar{R}$.

\begin{proposition}\label{t4}
	Consider the PWQ NN \eqref{nn2}. With a non-negative $r$ and a negative $b$, the function $\hat{J}(\cdot, \theta)$ that the NN produces is continuously differentiable and convex w.r.t. its input.
\end{proposition}

\textbf{Proof} In the interior of any $\hat{\mathcal{R}}_j, j=1, \ldots, \hat{N}_{\mathrm{r}}$, continuous differentiability of $\hat{J}(\cdot, \theta)$ is clear since $\hat{J}(\cdot, \theta)$ has a quadratic form, and convexity of $\hat{J}(\cdot, \theta)$ follows from the positive semi-definiteness of $\bar{R}$ in \eqref{e27}. Then, we have $\hat{P}_{j}-P^{*} \geq 0$ for all $j=1, \ldots, \hat{N}_{\mathrm{r}}$.

At the boundary of any neighboring $\hat{\mathcal{R}}_i$ and $\hat{\mathcal{R}}_j$, without loss of generality, suppose that $\hat{\mathcal{R}}_i$, $\hat{\mathcal{R}}_j$ are partitioned by the hyperplane $W_{1, \cdot}x+b_{1}=0$, i.e., $W_{1, \cdot} x+b_{1} \leq 0 \quad \forall x \in \hat{\mathcal{R}}_{i} \text { and } W_{1, \cdot} x+b_{1} \geq 0 \quad \forall x \in \hat{\mathcal{R}}_{j}$. It follows from \eqref{e26} that
\begin{equation}\label{e29}
	\hat{J}^j(x) = \hat{J}^i(x) + r_1(W_{1, \cdot} x + b_{1})^2
\end{equation}
Differentiating both sides of \eqref{e29} yields
\begin{equation*}
	\nabla_{x}\hat{J}_{j}(x,\theta)=\nabla_{x}\hat{J}_{i}(x,\theta) \quad \forall x \in \left\{ {x \in {\mathbb{R}^n}|{W_{1, \cdot}}x + {b_{1}} = 0} \right\}
\end{equation*}
which proves continuous differentiability of $\hat{J}(x, \theta)$ at the boundary. 

Furthermore, since $\hat{J}(x, \theta)$ is differentiable, convexity of $\hat{J}(x, \theta)$ at the boundary can be checked through first-order conditions \cite{boyd2004convex}. Formally speaking, $\hat{J}(x, \theta)$ is convex if and only if 
\begin{equation}\label{e31}
	\hat{J}(x_2, \theta) - \hat{J}(x_1, \theta) \geq \nabla^T_x \hat{J}(x_1, \theta)(x_2-x_1)
\end{equation}
holds for all $x_1,x_2$ belonging to the domain of $\hat{J}(x, \theta)$. As convexity of $\hat{J}(x, \theta)$ in the interior of any $\hat{\mathcal{R}}_j, j=1, \ldots, \hat{N}_{\mathrm{r}}$ has been verified, we only focus on the boundary. Without loss of generality, let $x_1 \in \hat{\mathcal{R}}_i$ and $x_2 \in \hat{\mathcal{R}}_j$. From \eqref{e29} one has
\begin{align*}
	&\hat{J}\left(x_{2}, \theta\right)-\hat{J}\left(x_{1}, \theta\right)-\nabla_{x} \hat{J}\left(x_{1}, \theta\right)^{T}\left(x_{2}-x_{1}\right)\nonumber\\
	=& \hat{J}_{i}\left( \!x_{2} \!\right)\!+\!r_{1}\left( \!W_{1, \cdot} x_{2}\!+\!b_{1} \!\right)^{2}\!-\!\hat{J}_{i}\left(\!x_{1}\!\right)\!-\!\nabla^T_{x} \hat{J}_{i}\left(x_{1},\theta\right)\left(\!x_{2}\!-\!x_{1}\!\right) \nonumber\\
	=& x_{2}^{T} P_{i} x_{2}+q^T_{i} x_{2}+r_{1}\left(W_{1, \cdot} x_{2}+b_{1}\right)^{2}\nonumber\\
	& -x_{1}^{T} P_{i} x_{1}+q^T_{i} x_{1}-\left(x_{1}^{T} P_{i}+q^T_{i}\right)\left(x_{2}-x_{1}\right) \nonumber\\
	=&\left(x_{2}\!-\!x_{1}\right)^{T} P_{i}\left(x_{2}\!-\!x_{1}\right)+r_{1}\left(W_{1, \cdot} x_{2}+b_{1}\right)^{2}\geq 0
\end{align*}
which demonstrates that $\hat{J}(\cdot, \theta)$ satisfies \eqref{e31} at the boundary. This completes the proof of Proposition \ref{t4}. \qed

\begin{remark}
	\textcolor{blue}{In \cite{chakrabarty2019approximate}, a quadratic value function approximator is proposed to implement both model-based and model-free ADP algorithms. Compared with the model-based method in \cite{chakrabarty2019approximate},} our method retains the PWQ property of the real value function and moves the training process offline. As a result, the proposed control framework can more accurately accommodate the optimal control law. Besides, \cite{chakrabarty2019approximate} assumes that the initial state lies in an invariant set, while our method relaxes this restriction. In particular, the domain of attraction under the proposed control law can be equal to the maximal stabilizable set $\bar{X}$. Overall, \textcolor{blue}{the benefit of considering piecewise quadratic approximation is its good approximation performance, while the advantage of considering a quadratic form is that one can use Riccati Equation to update the value function without any state transition data.}
\end{remark}

\subsection{Suboptimal control law based on dynamic programming}

With a well-fitted $\hat{J}(\cdot, \theta)$ available, at each time step $t\in \mathbb{N}$, we can obtain a suboptimal control policy by solving the following one-step dynamic programming problem
\begin{align}\label{onestep}
		&\mathop {\min }\limits_{{u_t}} \;\hat Q(x_t,u_t) \triangleq x_t^TQ{x_t} + u_t^TR{u_t} + \hat J\left( {A{x_t} + B{u_t},\theta} \right) \nonumber \\
		&{\text{s.}}{\text{t.  }} u_t \in \mathcal{U}, A{x_t} + B{u_t} \in \mathcal{C}
\end{align}
where $\hat Q(x_t,u_t)$ can be viewed as the approximated optimal Q-function \cite{bertsekas2019reinforcement} for problem \eqref{infinite0}. Denote the solution to \eqref{onestep} by $\hat{u}^*_t$. Here, we use another subscript $(\cdot)_t$ to indicate that problem \eqref{onestep} is solved online at each time step $t$. In problem \eqref{onestep}, $\mathcal{C}$ is chosen as $\bar{X}$ or $\mathbb{R}^n$, depending on whether $\bar{X}$ is computable. In particular, if the iterative algorithm (Algorithm 10.3) in \cite{borrelli2017predictive} does not terminate in finite time, we drop the constraint on $A{x_t} + B{u_t}$. This could happen, e.g., when there is no state constraint in \eqref{infinite0}. Besides, in both cases problem \eqref{onestep} is recursively feasible since $\bar{X}$ is a control invariant set (CIS) \cite{borrelli2017predictive}. Moreover, it should be noted that some approximation methods of predictive control laws specify $\mathcal{C}$ as the maximal CIS $\mathcal{C}_\infty$ \cite{chen2018approximating} or an arbitrary polytopic CIS \cite{karg2020efficient}, which may make the approximated controllers not stabilizing since usually $\bar{X} \subseteq \mathcal{C}_\infty$. The set $\mathcal{C}_\infty \backslash \bar{X} $ contains some initial states that cannot be steered to the origin. The dynamic programming solution in \eqref{onestep}, on the other hand, can prevent $x_t$ from falling into $\mathcal{C}_\infty \backslash \bar{X} $ as we can prove that $\hat{J}(\cdot,\theta)$ is possible to be a Lyapunov function for the closed-loop system. 

The following lemma characterizes the properties of the optimizer $\hat{u}^*_t(x_t)$ of \eqref{onestep}.

\begin{lemma}
	With Assumption A1 and the conditions in Proposition \label{t4} satisfied, the optimizer $\hat{u}^*_t(x_t)$ of \eqref{onestep} is a continuous PWA function on $\mathcal{C}$. 
\end{lemma}

\textbf{Proof} The proof is similar to standard arguments for proving the PWA property of explicit MPC law \cite[Theorem 17.1]{borrelli2017predictive}. In \eqref{onestep}, $\hat{J}\left(\cdot, \theta\right)$ has the PWQ form shown in \label{e26}. For a certain $\hat{\mathcal{R}}_j$, considering the following parametric quadratic program:
	\begin{align*}\label{convexqp}
		V_j (x_t)=&\mathop {\min }\limits_{{u_t}} x_t^TQ{x_t} + u_t^TR{u_t} + \hat J^j(Ax_t+Bu_t) \nonumber \\
		&{\text{s.}}{\text{t.  }} u_t \in \mathcal{U}, A{x_t} + B{u_t} \in \hat{\mathcal{R}}_j
	\end{align*}
	which admits a unique PWA optimizer \cite[Theorem 6.7]{borrelli2017predictive}
	\begin{equation*}
		u_t^j(x_t)=\tilde{F}^{j, k} x(0)+\tilde{g}^{j, k}, \quad \forall x(t) \in \mathcal{T}^{j, k}, \quad k=1, \ldots, \hat{N}_{\mathrm{r},j}
	\end{equation*}
	where $\{\mathcal{T}^{j, k}\}^{N_j}_{k=1}$ is a polyhedral partition of the polyhedral set $\mathcal{D}^j$ of feasible states for the above parametric quadratic program. The union of all $\mathcal{D}^j$ is thereby the feasible set $\mathcal{D}$ for problem in \eqref{onestep}. Also, $V_j (\cdot)$ is a PWQ function with the partition $\mathcal{T}^{j, k}, \quad k=1, \ldots, \hat{N}_{\mathrm{r},j}$. Then, we extend the domain $\mathcal{D}^j$ of each $V_j (\cdot)$ to $\mathcal{D}$ by assigning an infinite value for those $x_t \in \mathcal{D}\backslash \mathcal{D}^j$. As a result, the optimizer of problem \eqref{onestep} is determined by comparing the value of $V_j (x_t)$ and selecting the PWA feedback law $u_t^j\left(x_t\right)$ corresponding to the smallest $V_j (x_t)$:
	\begin{equation*}
		u_t^*(x_t) \!=  \!u_t^j(x_t), \mathrm{if} \;V_j (x_t) \;\mathrm{is \;the\; smallest \; among}\; \{V_l (x_t)\}^{\hat{N}_\mathrm{r}}_{l=1}
	\end{equation*}
	
	After this procedure, $u_t^*(\cdot)$ consists of affine functions. Each of affine function is defined on a region, which can have affine and quadratic boundaries. Quadratic boundaries may arise from the comparison of all $V_j (x_t), \;j=1,2,...,\hat{N}_{\mathrm{r}}$. However, it can be easily shown that problem \eqref{onestep} is strictly convex, because $\hat{J}\left(\cdot, \theta\right)$ is convex and $R>0$. Therefore, the optimizer $u_t^*(\cdot)$ is unique according to \cite[Proposition 1. 1. 2]{bertsekas1997nonlinear}. This implies that $u_t^j(x_t) = u_t^i(x_t)$ for those $x_t$ belonging to the quadratic boundary $V_j(x_t) = V_i(x_t)$. This can happen only if the quadratic boundary degenerates to a single feasible point or to affine boundaries. Therefore, $u_t^*(\cdot)$ is a PWA function. 
	
	Finally, note that the feasible set for $u_t$ at any $x_t \in \mathcal{C}$ is a continuous point-to-set map on $\mathcal{C}$. The continuity of $u_t^*(\cdot)$ follows from the uniqueness of $u_t^*(\cdot)$, the continuity of $\hat{Q}(\cdot,\cdot)$, and \cite[Corollary 8.1]{hogan1973point}. \qed


The objective function in \eqref{onestep} is nonlinear and contains a neural network. Standard solvers such as the ellipsoid algorithm or the interior-point algorithm require the computation of the Hessian $\nabla^{2}_u \hat Q(x,u)$ or the gradient $\nabla_u \hat Q(x,u)$ in each iteration. Such computation can only be carried out by visiting all units in the hidden layers and extracting the activated ones. The advantage of low computational complexity brought by dynamic programming will then inevitably diminish.

In view of this, we intend to avoid frequently calculating $\nabla^{2}_u \hat Q(x,u)$ and $\nabla_u \hat Q(x,u)$ by decomposing the Q-function $\hat Q(x_t,u_t)$ into some quadratic functions. In particular, we develop two optimization algorithms in which problem \eqref{onestep} is reduced to a QP problem in each iteration. For a given $x_t$, $t \in \mathbb{N}$, consider the set of activated ReLU units in $\hat{J}\left(A x+B u, \theta\right)$:
\begin{equation}\label{active2}
	\mathcal{\bar A}(u)=\left\{i\in \{1,\dots,M\}\;|\; W_{i, \cdot} (Ax_t+Bu)+b_{i}>0\right\}
\end{equation}
where $M$ is the width of the hidden layer. With \eqref{active2}, $\hat Q(x_t,u)$ can thus be computed as
\begin{equation}\label{e40}
	\hat Q(x_t,u) = u^{T} \bar{P}(\mathcal{\bar A}(u)) u + \bar {q}^T(\mathcal{\bar A}(u))u +\bar{v}(\mathcal{\bar A}(u)) 
\end{equation}
where
\vspace{-20pt}
\begin{small}
	\begin{equation}\label{e41}
	\begin{aligned}
		&\bar{P}(\mathcal{\bar A}(u))=R+B^{T}(P^*+ \sum_{i \in \bar{\mathcal{A}}(u)} r_{i} W_{i, \cdot}^{T} W_{i, \cdot}) B \\
		&\bar {q}^T(\mathcal{\bar A}(u))= 2x^T_tA^TP^*B+ 2(\sum_{i \in \bar{\mathcal{A}}(u)} r_{i}\left(W_{i, \cdot} A x_t+b_{i}\right) W_{i, \cdot}) B \\
		&\bar{v}(\mathcal{\bar A}(u))=x_{t}^{T} (Q+A^TP^*A) x_{t}+\sum_{i \in \bar{\mathcal{A}}(u)} r_{i}\left(W_{i, \cdot} A x_t+b_{i}\right)^{2}
	\end{aligned}
\end{equation}\end{small}Since $\bar{P}(\mathcal{\bar A}(u))>0$, the right-hand side of \eqref{e40} is a convex quadratic function if $\bar{\mathcal{A}}(u)$ is fixed. An algorithm that can cope with general piecewise convex programs (PCP) is proposed in \cite{louveaux1978piecewise}. We adapt it to solving problem \eqref{onestep} and summarize it in Algorithm 1.

\begin{algorithm}
	\caption{PCP algorithm for solving the piecewise quadratic program \eqref{onestep} \cite{louveaux1978piecewise}}
	\label{alg:A}
	\begin{small}
	\begin{algorithmic}
		\State \textbf{Input:} State $x_t$ at time step $t$, input $u_{t-1}$ at the last time step $t-1$ (if $t>0$), the optimal Q-function $\hat{Q}\left(\cdot, \cdot\right)$, $\mathcal{C}$
		
		\State \textbf{Output:} Control input $u_t$ that will be applied to the system
		\State
		
		\State Initialize a starting point $u^{(1)} \leftarrow u_{t-1}$
		\State Initialize a set $\mathcal{U}_1 \leftarrow\{u|u \in \mathcal{U}, A x_{t}+B u \in \mathcal{C}$
		\For {$s = 1,2,\dots$} 
		\State Update the set of activated units $\bar{\mathcal{A}}(u^{(s)})$ at $u^{(s)}$ by \eqref{active2}, and compute the coefficients $\bar{P}(\mathcal{\bar A}(u^{(s)}))$ and $\bar {q}^T(\mathcal{\bar A}(u^{(s)}))$ associated with $\bar{\mathcal{A}}(u^{(s)})$ through \eqref{e41}
		\State Find $u^{(s+1)}\leftarrow\mathop {\arg\min }\limits_{{u}\in \mathcal{U}_s} \;u^{T} \bar{P}(\mathcal{\bar A}(u^{(s)})) u+\bar {q}^T(\mathcal{\bar A}(u^{(s)})) u $
		\State Find $d^{(s+1)}\leftarrow\mathop {\arg\min }\limits_{{u}} \;u^{T} \bar{P}(\mathcal{\bar A}(u^{(s)})) u+\bar {q}^T(\mathcal{\bar A}(u^{(s)})) u$
		\vspace{-10pt}
		\begin{equation*}\label{algo2}
			\begin{aligned}
				{\text{\;\;\;\;\;\;\;\;\;\;\;\;\;\;\;\;\;\;s.}}{\text{t.}}\;\; &W_{i, \cdot}(Ax_t+Bu)+b_{i}\geq 0  \quad \forall i \in \bar{\mathcal{A}}(u^{(s)})\\ 
				&W_{j, \cdot}(Ax_t+Bu)+b_{j}\leq 0  \quad \forall j  \notin \bar{\mathcal{A}}(u^{(s)}) \\
				&u \in \mathcal{U}, A{x_t} + B{u} \in \mathcal{C}
			\end{aligned} 
		\end{equation*} 
		\If{$|d^{(s+1)}-u^{(s+1)}| \leq \varepsilon $}
		\State Let $s_\text{m} \leftarrow s+1$, \textbf{return} $u_t \leftarrow d^{(s_\text{m})}$, and \textbf{break}
		\Else
		\State Let $\mathcal{U}_{s+1} \leftarrow \mathcal{U}_s \cap \{u|(u^{(s+1)T}\bar{P}(\mathcal{\bar A}(u^{(s)}))+\bar {q}^T(\mathcal{\bar A}(u^{(s)})))(u-d^{(s+1)})\leq 0 \}$
		\EndIf
		\EndFor
	\end{algorithmic}
\end{small}
\end{algorithm}

Algorithm 1 is a modified version of the PCP algorithm in \cite{louveaux1978piecewise}. In each iteration, Algorithm 1 solves two auxiliary QPs and terminates if the minimizers $d^{(s+1)}$ and $u^{(s+1)}$ are identical. The parameter $\varepsilon >0$ is a tolerance. If $|d^{(s+1)}-u^{(s+1)}| \leq \varepsilon $ is recorded, we assume that the two problems in Algorithm 1 admit the same optimal point. It is preferable that $d^{(s_\text{m})}$ or $u^{(s_\text{m})}$ could be the exact solution to problem \eqref{onestep} and that Algorithm 1 could stop in finite time. \cite{louveaux1978piecewise} provides these guarantees, which are summarized in the following theorem.
\begin{thm}[\cite{louveaux1978piecewise}]\label{t50}
	Consider the dynamic programming problem \eqref{onestep}. For any $x_t$ that makes problem \eqref{onestep} feasible, the PCP algorithm (Algorithm 1) terminates in a finite number of iterations, i.e., there exists a finite $s_\mathrm{m}$ such that $|d^{(s_\mathrm{m})}-u^{(s_\mathrm{m})}| \leq \varepsilon$ holds. Moreover, if $|d^{(s_\mathrm{m})}-u^{(s_\mathrm{m})}| \leq \varepsilon$ holds, then $d^{(s_\text{m})}$ is the solution to problem \eqref{onestep}.
\end{thm}

As shown in \cite{louveaux1978piecewise}, using Algorithm 1 to solve piecewise quadratic programs can be effective since it only needs to compute $\nabla^{2}_u \hat Q(x,u)$ or $\nabla_u \hat Q(x,u)$ when $s$ is updated. On the other hand, Algorithm 1 is not the ideal choice for solving our DP problem, because problem \eqref{algo2} contains numerous constraints if the neural network has a large number of hidden units. This motivates us to consider the following design. In each iteration $s$, $s\in \mathbb{N}^+$, let $u^{(s)}$ be an initial input (for $s=1$) or the input calculated from the last iteration (for $s>1$). We compute the set of activated units $\bar{\mathcal{A}}(u^{(s)})$ at $u^{(s)}$, and thereby get $\bar{P}(\mathcal{\bar A}(u^{(s)}))$ and $\bar {q}^T(\mathcal{\bar A}(u^{(s)}))$ from \eqref{e41}. Then, we solve the following QP:
\begin{equation}\label{e42}
	\begin{gathered}
		u^{(s+1)}=\mathop {\arg\min }\limits_{{u}} \;u^{T} \bar{P}(\mathcal{\bar A}(u^{(s)})) u+\bar {q}^T(\mathcal{\bar A}(u^{(s)})) u \hfill \\
		{\text{s.}}{\text{t.  }} u \in \mathcal{U}, A{x_t} + B{u} \in \mathcal{C}\hfill \\ 
	\end{gathered} 
\end{equation} 
which returns $u^{(s+1)}$ for the next iteration. In the next iteration, after computing $\bar{\mathcal{A}}(u^{(s+1)})$, we can terminate and output $u^{(s+1)}$ if $\bar{\mathcal{A}}(u^{(s+1)})$=$\bar{\mathcal{A}}(u^{(s)})$. If a cycle occurs, i.e., $\bar{\mathcal{A}}(u^{(s+1)})$=$\bar{\mathcal{A}}(u^{(k)}),\;\exists k \in \{1,\dots,s-1\}$, we arbitrarily choose another $\bar{\mathcal{A}}(u^{(s+1)})$ that has not been involved in previous iterations. The proposed method for getting the solution to \eqref{onestep} is summarized in Algorithm 2. 

\begin{algorithm}
	\caption{Decomposition algorithm for solving the piecewise quadratic program \eqref{onestep}}
	\label{alg:B}
	\begin{algorithmic}
		\State \textbf{Input:} State $x_t$ at time step $t$, input $u_{t-1}$ at the last time step $t-1$ (if $t>0$), the optimal Q-function $\hat{Q}\left(\cdot, \cdot\right)$, $\mathcal{C}$
		
		\State \textbf{Output:} Control input $u_t$ that will be applied to the system
		\State
		
		\State Initialize a starting point $u^{(1)} \leftarrow u_{t-1}$
		\For {$s = 1,2,\dots$} 
		\State Update the set of activated units $\bar{\mathcal{A}}(u^{(s)})$ by \eqref{active2}
		
		\If {$\bar{\mathcal{A}}(u^{(s)}) = \bar{\mathcal{A}}(u^{(s-1)})$ and $ s>1$}
		
		\State Let $s_\mathrm{m}\leftarrow s$, \textbf{return} $u_t\leftarrow  u^{(s_\mathrm{m})}$, \textbf{break}
		
		\Else
		\If{$\bar{\mathcal{A}}(u^{(s)}) = \bar{\mathcal{A}}(u^{(k)}), \exists k \in \{1,\dots,s-2\}$ and $s>2$}
		\State Reset $\bar{\mathcal{A}}(u^{(s)})$ to be a new set of activated units that never occurred previously.
		\EndIf
		\State Compute the coefficients $\bar{P}(\mathcal{\bar A}(u^{(s)}))$ and $\bar {q}^T(\mathcal{\bar A}(u^{(s)}))$ associated with $\bar{\mathcal{A}}(u^{(s)})$ through \eqref{e41}
		\State Update the policy through \eqref{e42} and get $u^{(s+1)}$		
		\EndIf
		\EndFor
	\end{algorithmic}
\end{algorithm}
In Algorithm 2, the starting point $u^{(1)}=u_{t-1}$ is initialized with the last control input, which can be viewed as a warm start for the algorithm. To understand the rationale of this design, suppose that $x_t$ is close to the origin after enough time steps. Then, $u_t$ will vary slightly around the origin. Choosing $u^{(1)}=u_{t-1}$ is beneficial to reducing the number of iterations.

Compared to Algorithm 1, Algorithm 2 only needs to solve one QP, in which the constraints are the same as those in \eqref{onestep}. Additionally, Algorithm 2 circumvents the calculation of $\mathcal{U}_{s+1}$.  Meanwhile, Algorithm 2 can also achieve finite termination as well as the optimality for problem \eqref{onestep}. However, the number of iterations in Algorithm 2 may be more than that in Algorithm 1, and we need to store all the previous set of activated units $\bar{\mathcal{A}}(u^{(k)}),\;k=1,\dots,s-1$.  For all that, we can still rely on Algorithm 2 because we have found that cycles rarely occur in our numerical experiments for several examples. Besides, we can also switch to algorithm 1 by taking $\mathcal{U}_{s_\mathrm{c}} \leftarrow\{u|u \in \mathcal{U}, A x_{t}+B u \in \mathcal{C}$ if a cycle is detected at $s = {s_\mathrm{c}}$.

\begin{thm}\label{t5}
	Consider the DP problem \eqref{onestep}. For any $x_t$ that makes problem \eqref{onestep} feasible, the decomposition algorithm (Algorithm 2) terminates in a finite number of iterations, i.e, there exists a finite $s_\mathrm{m}$ such that $\bar{\mathcal{A}}\left(u^{(s_\mathrm{m})}\right) = \bar{\mathcal{A}}\left(u^{(s_\mathrm{m}-1)}\right)$ holds. Moreover, if $\bar{\mathcal{A}}\left(u^{(s_\mathrm{m})}\right) = \bar{\mathcal{A}}\left(u^{(s_\mathrm{m}-1)}\right)$, then $u^{(s_\mathrm{m})}$ is the solution to problem \eqref{onestep}.
\end{thm}

\textbf{Proof} We will first show that if Algorithm 2 stops, it outputs a solution to problem \eqref{onestep}.

According to Algorithm 2, $u^{(s_\text{m})}$ minimizes $\bar{Q}_{u^{(s_\text{m})}}\left(x_{t}, u\right)\triangleq u^{T} \bar{P}(\mathcal{\bar A}(u^{(s_\text{m})})) u+\bar {q}^T(\mathcal{\bar A}(u^{(s_\text{m})})) u+\bar{v}(\mathcal{\bar A}(u^{(s_\text{m})}))$ subject to $u \in \mathcal{U}, A x_{t}+B u \in \mathcal{C}_{\infty}$ if $\bar{\mathcal{A}}(u^{(s_\text{m})}) = \bar{\mathcal{A}}(u^{(s_\text{m}-1)})$. In this case, the inequality
\begin{equation}\label{jj}
	\nabla^T_u\bar{Q}_{u^{(s_\text{m})}}( x_t,u^{(s_\text{m})}) (u-u^{(s_\text{m})}) \geq 0
\end{equation}
holds for all $u \in \mathcal{U}_1 \triangleq \{u|u \in \mathcal{U}, A x_{t}+B u \in \mathcal{C}$. Using the fact that $\nabla^T_u\bar{Q}_{u^{(s_\text{m})}}\left(x_t,u^{(s_\text{m})}\right) = \nabla^T_u\hat{Q}\left(x_{t}, u^{(s_\text{m})}\right)$, the gradient inequality for the convex PWQ function $\hat{Q}\left(x_{t}, \cdot\right)$ can be applied at $u^{(s_\text{m})}$:
\begin{equation*}\label{e44}
	\hat{Q}(x_{t}, u) \geq \hat{Q}(x_{t}, u^{(s_\text{m})})+ \nabla^T_u\hat{Q}_{u^{(s_\text{m})}}(x_t, u^{(s_\text{m})})(u-u^{(s_\text{m})})
\end{equation*}
which, combined with inequality \eqref{jj}, shows that $\hat{Q}\left(x_{t}, u\right) \geq \hat{Q}\left(x_{t}, u^{(s_\text{m})}\right)$ for all $u \in \mathcal{U}_1$. Thus, the optimality of $u^{(s_\text{m})}$ is proven.

Furthermore, if a cycle occurs, i.e., $\bar{\mathcal{A}}(u^{(s)}) = \bar{\mathcal{A}}(u^{(k)}), \exists k \in \{1,2,\dots,s-2\}$ happens for some $s$, according to Algorthm 2, we select another set of activated units that has never been considered in problem \eqref{e42}. As the number of combinations of activated units is finite, which means the number of different $\bar{\mathcal{A}}(\cdot)$ is limited, the algorithm must stop in a finite number of iterations. \qed

\textcolor{blue}{In general, the amount of data needed in our method is in general less than that in policy-based methods \cite{karg2020efficient} because the value function is a scalar, but the policy may be multi-dimensional. Besides, for the extension to constrained nonlinear systems, there are some situations when the value function can be guaranteed to be continuous \cite[Theorem 3]{postoyan2016stability}, while ensuring the continuity of the optimal policy is almost impossible. Basically, learning a continuous function requires simpler approximation structures than learning a discontinuous function.}

\subsection{Analysis of the proposed method}
\subsection{Stability analysis}
The recursive feasibility of the DP problem \eqref{onestep} is inherently guaranteed. In this section, we investigate the stability of the proposed control law.

If problem \eqref{empc} is non-degenerate, all $\mathcal{R}_{i},\;i=1,\dots,N_{\mathrm{r}}$ are full-dimensional \cite{borrelli2017predictive}. In contrast, in the degenerate case, some $\mathcal{R}_{i}$ are lower-dimensional, and in general, they correspond to common boundaries between full-dimensional regions. Since both $J_{\infty}^{*}(\cdot)$ and $\hat{J}\left(\cdot, \theta\right)$ are PWQ on polyhedra, the intersection of any $\mathcal{R}_{i}$ and $\hat{\mathcal{R}}_{j},\;i=1, \ldots, N_{\mathrm{r}},\;j=1, \ldots, \hat{N}_{\mathrm{r}}$ is still polyhedral. Let ${\mathcal{R}}_{i,j}$ denote this intersection if such intersection represents a full-dimensional region, i.e., ${\mathcal{R}_{i,j}} \triangleq {\mathcal{R}_i} \cap {\hat{\mathcal{R}}_j},\;i \in \left\{ {1,...,{N_\mathrm{r}}} \right\}\;,\;j \in \{{1,...,{{\hat N}_\mathrm{r}}} \}\;{\text{and}}\;\mathrm{dim}({\mathcal{R}_i} \cap {\hat{\mathcal{R}}_j}) = n$. It is clear that all ${\mathcal{R}_{i,j}}$ are a partition of $X_0$.

\textbf{Upper bound of the approximation error.} To certificate the stability, we need to know the upper bound of the approximation error for all $x$ belonging to $X_0$. Computing the maximum of ${| {\hat J( \cdot ,\theta ) - J_\infty ^*( \cdot )} |/J_\infty ^*( \cdot )}$ over each ${\mathcal{R}_{i,j}}$ is cumbersome since (i) it needs the analytical form of $J_\infty ^*( \cdot )$ and (ii) it needs to solve at most $N_\mathrm{r} \hat N_{\mathrm{r}}$ non-convex optimization problems, which is computationally demanding. 

For this reason, we develop a procedure to get an upper bound of ${| {\hat J( \cdot ,\theta ) - J_\infty ^*( \cdot )} |/J_\infty ^*( \cdot )}$ by leveraging the Lipschitz continuity of $\nabla \hat J( \cdot ,\theta )$ and $ \nabla J_\infty ^*( \cdot )$. Since we can compute the approximation error of the proposed neural network at the training points, it is possible to obtain an upper bound of the approximation error for all $x$ in $X_0$. In the interior of each ${\mathcal{R}_{i,j}}$, both $J_{\infty}^{*}(\cdot) $ and $\hat{J}\left(\cdot, \theta\right)$ are quadratic and hence twice continuously differentiable. According to \cite[Lemma 3.2]{borrelli2017predictive}, the gradients ${\nabla _x}J_{\infty}^{*}(\cdot) $ and ${\nabla _x}\hat{J}\left(\cdot, \theta\right)$ are locally Lipschitz on $\mathrm{int}(\mathcal{R}_{i,j})$, i.e., there exist non-negative constants $L_i$ and $\hat{L}_j$ such that for all pairs $(x,y) \in \mathrm{int}(\mathcal{R}_{i,j}) \times \mathrm{int}(\mathcal{R}_{i,j})$, we have
\begin{equation}\label{Lipschitz}
	\begin{aligned}
		&J_\infty ^*\left( y \right) \leq J_\infty ^*\left( x \right) + {\nabla ^T}J_\infty ^*\left( x \right)(y - x) + \frac{{{L_i}}}{2}\left\| {y - x} \right\|_2^2\\
		&\hat{J}\left(y, \theta \right) \leq \hat{J}\left(x, \theta \right) + {\nabla ^T}\hat{J}\left(x, \theta \right)(y - x) + \frac{{{\hat{L}_j}}}{2}\left\| {y - x} \right\|_2^2\\
	\end{aligned}
\end{equation}
The Lipschitz constants $L_i$ and $\hat{L}_j$ can be chosen as the largest eigenvalue of $P_{i}$ and $\hat{P}_{j}$ \cite[Lemma 3.2]{borrelli2017predictive}, respectively. In the following, we will show that there exists a positive constant $\zeta$ such that 
\begin{equation}\label{error2}
	|e(x)| \triangleq \left| {\frac{{\hat J\left( {x,\theta } \right)}}{{J_\infty ^*\left( x \right)}} - 1} \right| \leq {\zeta },\quad \forall x \in  X_0
\end{equation}
To achieve this, we need the following assumption.

\textbf{Assumption A2:} For the partition $\mathcal{R}_{i,j}$ of $X_0$, suppose that there exists at least one training point in each $\mathrm{int}(\mathcal{R}_{i,j})$. 

Meanwhile, define a measure of the gradient error as $e_\mathrm{grad}\left( x \right) \triangleq \frac{{{{\left\| {\nabla \hat J\left( {x,\theta } \right) - \nabla J_\infty ^*\left( x \right)} \right\|}_2}}}{{J_\infty ^*\left( x \right)}}$, and let $\bar{e}$ and $\bar{e}_\mathrm{grad}$ stand for the maximum values of $|e(\cdot)|$ and $e_\mathrm{grad}(\cdot)$ over all training data, which are computable. It is straightforward to see that
\begin{equation}\label{assumption2}
	|e(x_{k}^{(i*)})|\! \leq\! \bar{e},\; e_\mathrm{grad}(x_{k}^{(i*)}) \!\leq\! \bar{e}_\mathrm{grad},\;i\!=\!1, \ldots, N_{x},\;k\!=\!0, \ldots, N-1
\end{equation}
where $x_{k}^{(i*)}$ is any sampled state specified in \eqref{sufficient}.

\begin{remark}
	\textcolor{blue}{Assumption A2 is used to compute a global bound $\zeta$ to guarantee \eqref{error2}. Assumption A2 is not required to satisfy when training the NN. To guarantee stability, if Assumption A2 is not satisfied after the NN is trained, we can add one testing point in each region $\mathrm{int}(\mathcal{R}_{i,j})$ where there is no training point. Then, we evaluate the approximation errors $|e(\cdot)|$ and $e_\mathrm{grad}(\cdot)$ at these testing points, and adjust $\bar{e}$ and $\bar{e}_\mathrm{grad}$ if necessary so that $\left|e\left(\cdot\right)\right| \leq \bar{e}, e_\mathrm{grad}\left(\cdot\right) \leq \bar{e}_\mathrm{grad}$ also hold at these testing points. Nevertheless, we should point out that one of the limitations is that we sometimes need carefully selected testing data to satisfy Assumption A2, and the data should be enough dense if the NN or the optimal value function has many polyhedral regions. To compute $\zeta$, one needs to compute (i) the partition for the value functions, (ii) the Lipschitz constants of the value functions, and (iii) $J_{\infty}^*$ at the sampled points. Relaxing Assumption A2 and exploring a more convenient way to compute the error bound  $\zeta$ will be our future work.}
\end{remark}

Let $\mathcal{R}_{1}$ refer to the polyhedron where no constraints in \eqref{empc} are active, i.e., $\mathcal{R}_{1} = \mathcal{O}^\text{LQR}_{\infty}$, and accordingly let $\hat{\mathcal{R}}_{1}$ represents the polyhedron where no ReLU units in $\hat{J}\left(\cdot, \theta \right)$ are activated. In the region $\mathcal{R}_{1,1}=   \mathcal{R}_{1} \cap \hat{\mathcal{R}}_{1}$, we have $|e(x)| \equiv 0$ due to \eqref{nn2}. For every $\mathcal{R}_{i,j}$ except $\mathcal{R}_{1,1}$, consider the following three cases:

\textbf{Case 1:} $e\left( x \right) \geq 0,\;\forall x \in \mathrm{int}(\mathcal{R}_{i,j})$. In this case, for any $x \in \mathrm{int}(\mathcal{R}_{i,j})$, let $y \in \mathrm{int}(\mathcal{R}_{i,j})$ denote the training or testing point closest  to $x$. Substituting $x$ and $y$ into \eqref{Lipschitz} results in
\vspace{-20pt}
\begin{small}
\begin{align}\label{check1}
		&\left| {e\left( x \right)} \right| = \frac{{\hat J\left( {x,\theta } \right)}}{{J_\infty ^*\left( x \right)}} - 1 \nonumber\\
		\leq& \frac{{\hat J\left( {{y},\theta } \right) + {\nabla ^T}\hat J\left( {{y},\theta } \right)\left( {x - {y}} \right) + {{\hat L}_j}\left\| {x - {y}} \right\|_2^2/2}}{{J_\infty ^*\left( {{y}} \right) + {\nabla ^T}J_\infty ^*\left( {{y}} \right)\left( {x - {y}} \right)}} - 1  \nonumber\\
		 \leq&\! \frac{{\hat J\!\left( {{y},\theta } \right) \!-\! J_\infty ^*\left( {{y}} \right) \!+ \!{{( {\nabla \!\hat J\left( {{y},\theta } \right) \!- \!\nabla \!J_\infty ^*\left( {{y}} \right)} )}^T}\!\left( {x\! -\! {y}} \right) \!+ \!{{\hat L}_j}\!\left\| \!{x\! -\! {y}} \right\|_2^2/2}}{{J_\infty ^*\left( {{y}} \right) \!-\! {{\left\| {{\nabla ^T}J_\infty ^*\left( {{y}} \right)} \right\|}_2}\left\| {x - {y}} \right\|_2^{}}}\nonumber \\ 
		 \leq& \frac{{\left| {e\left( {{y}} \right)} \right| +  {e_\mathrm{grad}\left( {{y}} \right)} \left\| {x - {y}} \right\|_2^{} + {{\hat L}_j}\left\| {x - {y}} \right\|_2^2/\left( {2J_\infty ^*\left( {{y}} \right)} \right)} }{{1 - {\beta(y)}\left\| {x - {y}} \right\|_2^{}}}
\end{align}
\end{small}where ${{\beta(y)}} = {{{\left\| {{\nabla ^T}J_\infty ^*\left( {{y}} \right)} \right\|}_2}/J_\infty ^*\left( {{y}} \right)}$. The first inequality of \eqref{check1} is true due to the second inequality of \eqref{Lipschitz} and the convexity of $J_\infty ^*\left( \cdot \right)$. Let $d_{i,j}$ denote the maximum Euclidean distance between $x$ and its nearest training or testing point subject to $x \in \mathrm{int}(\mathcal{R}_{i,j})$. Using the fact that \eqref{assumption2} holds at $y$, we obtain from \eqref{check1} that
\begin{equation}\label{check2}
	\left| {e\left( x \right)} \right| \leq \frac{1}{{1 - {\beta (y)}d_{i,j}}}\left( {\bar e + \bar e_\mathrm{grad} d_{i,j} + \frac{{{{\hat L}_j}{{d_{i,j}}^2}}}{{2J_\infty ^*\left( {{y}} \right)}}} \right)
\end{equation}
holds for any $x \in \mathrm{int}(\mathcal{R}_{i,j})$. Herein, it should be mentioned that $\beta(y)$ is bounded on all $\mathcal{R}_{i,j}$ except $\mathcal{R}_{1,1}$, since $J_{\infty}^{*}\left(\cdot\right)$ can only equal zero at origin, which is contained in $\mathcal{R}_{1,1}$. Therefore, one can always make $d_{i,j}$ sufficiently small so that ${1-\beta(y) d_{i,j}} > 0$.

\textbf{Case 2:} $e\left( x \right) \leq 0,\;\forall x \in \mathrm{int}(\mathcal{R}_{i,j})$. Similarly to Case 1, we can readily get
\vspace{-20pt}
\begin{small}
\begin{align}\label{case2}
		&\left| {e\left( x \right)} \right| \!=\! 1\!-\!\frac{{\hat J\left( {x,\theta } \right)}}{{J_\infty ^*\left( x \right)}}\!\leq\! 1\!-\!\frac{{\hat J\left( {{y},\theta } \right) + {\nabla ^T}\hat J\left( {{y},\theta } \right)\left( {x - {y}} \right) }}{J_\infty ^*\left( {{y}} \right) + {\nabla ^T}J_\infty ^*\left( {{y}} \right)\left( {x - {y}} \right) +  \frac{L_i d_{i,j}^2}{2}}  \nonumber \\
		& \leq \frac{J_\infty ^*\left( {{y}} \right)\!-\!{\hat J\left( {{y},\theta } \right) \!+\! {{( { \nabla J_\infty ^*\left( {{y}} \right)} \!-\!\nabla \hat J\left( {{y},\theta } \right) )}^T}\left( {x \!- \!{y}} \right)\! + \!{ L_i}d_{i,j}^2/2}}{{J_\infty ^*\left( {{y}} \right) - {{\left\| {{\nabla ^T}J_\infty ^*\left( {{y}} \right)} \right\|}_2}\left\| {x - {y}} \right\|_2^{}}}\nonumber \\ 
		& \leq \frac{1}{{1 - {\beta(y)}d_{i,j}}}\left( {\bar e + \bar e_\mathrm{grad} d_{i,j} + \frac{{{{ L}_i}{{d_{i,j}}^2}}}{{2J_\infty ^*\left( {{y}} \right)}}} \right)
\end{align}
\end{small}where the first inequality of \eqref{case2} comes from the first inequality of \eqref{Lipschitz} and the convexity of $\hat{J}\left(\cdot, \theta\right)$. The last line of \eqref{case2} is almost the same as the right-hand side of \eqref{check2}, and the only difference is that $\hat{L}_{j}$ is replaced by $L_i$.


\textbf{Case 3:} $\exists x_1, x_2 \in  \mathrm{int}(\mathcal{R}_{i,j})$ such that $e(x_1)>0$ and $e(x_2)<0$. In this case, for all $x \in \mathrm{int}(\mathcal{R}_{i,j})$ subject to ${\hat J\left( {x,\theta } \right)} \geq {J_\infty ^*\left( x \right)}$, we can get the same upper bound for $|e(x)|$ as in \eqref{check2}, and for all $x \in \mathrm{int}(\mathcal{R}_{i,j})$ subject to ${\hat J\left( {x,\theta } \right)} < {J_\infty ^*\left( x \right)}$, inequalities \eqref{case2} holds. 

Consequently, combining the above 3 cases, it is sufficient to conclude that for any $x \in \mathrm{int}(\mathcal{R}_{i,j})$, $\left| {e\left( x \right)} \right|$ is upper bounded by 
\begin{equation}\label{bound}
	\zeta_{i,j} = \frac{1}{{1 - {\beta(y)}d_{i,j}}}\left( {\bar e + \bar e_\mathrm{grad} d_{i,j} + \frac{{{{(\hat L}_j+ L_i)}{d_{i,j}^2}}}{{2J_\infty ^*\left( {{y}} \right)}}} \right)
\end{equation}
Finally, since $e(x)$ is continuous on the closure of $\mathcal{R}_{i,j}$, the bound $\zeta_{i,j}$ applies to all $x$ in $\mathcal{R}_{i,j}$. In addition, thanks to Assumption A2, the value of $\zeta$ in \eqref{error2} can be determined by computing the right-hand side of \eqref{bound} at all training/testing points and choosing the largest one.


\textbf{Computation of $\bm{\nabla \hat J\left( {{y},\theta } \right)}$, $ \bm{\nabla J_\infty ^*\left( {{y}} \right)}$, $\bm{L_i}$, and $\bm{\hat{L}_j}$.} The computation of $\nabla \hat J\left( {{y},\theta } \right)$ and $\hat{L}_j$ is straightforward since the analytical form of $\hat J\left( {\cdot,\theta } \right)$ is known. The values of $\nabla J_\infty ^*\left( {y}\right)$ and $L_i$ can be collected in the training process. Specifically, note that all $y$ are contained in the interior of full-dimensional $\mathcal{R}_i$, in which the rows of $G_{\mathcal{A}^i}$ are linearly independent. In this case, the Lagrange multipliers $\lambda^*(\cdot)$ of problem \eqref{empc} are affine functions of $x$ \cite{bemporad2002explicit} on and thus continuously differentiable. Then, similar to the analysis in \cite[Theorem 6.9]{borrelli2017predictive}, the gradient $\nabla J_\infty ^*\left( {{y}} \right)$ can be computed by 
\begin{equation*}\label{gradient_J}
	\nabla J_\infty ^*\left( {{y}} \right) = -S^T\lambda^*(y)+ Y y + F U^*(y)
\end{equation*}
To obtain $L_i$, let us restrict our attention to the set of active constraints at $y$. The set $\mathcal{A}_i$ defined in Theorem \ref{t3} can be determined by checking whether the elements of $\lambda^*(y)$ is zero or not. Then, as illustrated in the proof of \cite[Lemma 1 (Theorem \ref{t3} in this paper)]{baotic2008efficient}, substituting $\mathcal{A}_i$ into
\begin{equation*}\label{Pj}
	P_i = P^*+\frac{1}{2}S_{\mathcal{A}_{i}, \cdot}^{T} \Gamma^{-1} S_{\mathcal{A}_{i}, \cdot}
\end{equation*}
where $\Gamma=G_{\mathcal{A}_{i}, \cdot} H^{-1} G_{\mathcal{A}_{i}, \cdot}^{T} \succ 0$, we can obtain $P_{i}$ and hence select $L_i$ as the largest eigenvalue of $P_{i}$.

With the property of boundedness for $e(x)$ established, we can assess the stability for the closed-loop system with the approximate controller $\hat{u}^*$. Corresponding to the selection of $\mathcal{C}$ in problem \eqref{onestep}, we consider two cases: (1)  $\mathcal{C} = \bar{X}$ and (2) $\mathcal{C} = \mathbb{R}^n$.

\begin{thm}\label{t6}
	Let $\hat{u}_{t}^{*}$ be the solution to problem \eqref{onestep} with $\mathcal{C} = \bar{X}$. Then, $\hat{u}_{t}^{*}$ is recursively feasible for the initial condition $ x_0 \in \bar{X} $. Furthermore, suppose that Assumptions A1-A2 hold with $X_0 = \bar{X}$. If $\zeta$ \eqref{error2} satisfies
	\begin{equation}\label{rho1}
		\frac{1-\zeta^2}{\zeta } > 2\mathop {\sup }\limits_{x \in X_0\backslash \left\{ 0 \right\}} \frac{{\hat J\left( {x,\theta } \right)}}{{{x^T}Qx}} ,
	\end{equation}
	then the origin of the closed-loop system $x_{t+1}=Ax_t+B\hat{u}_{t}^{*}, t=0,1,\dots$ is asymptotically stable with domain of attraction $ \bar{X} $.
\end{thm}

\textbf{Proof} \textcolor{blue}{In Theorem \ref{t6}, $\mathcal{C}=\bar{X}$. Since $\bar{X}$ is control invariant, recursive feasibility regarding both state and input constraints
	is guaranteed for all initial states in $\bar{X}$.} 

Let us first consider the solution to the MPC problem \eqref{e5}. For every $x_t, t=0,1,\dots$, MPC finds the solution $U_{t}^{*}$ to problem \eqref{e5} and only collects the first $m$ elements, i.e., $u_t^* = [{I_m}\;\;\;\underbrace {{0_m}\;\;\; \cdots \;\;\;{0_m}}_{N - 1\;{\text{terms}}}]U^{*} (x_t)$ will be applied to the system. From the equivalence between problem \eqref{infinite0} and problem \eqref{e5}, $\left\{u_t^*\right\}_{t=0}^\infty$ is also a solution to the problem \eqref{infinite0}. As a result, $Ax_t+Bu_t^* \in \bar{X}$ for any $x_t \in \bar{X}$. (Otherwise, $J_{\infty}^{*}(Ax_t+Bu_t^*)=\infty$ and then $J_{\infty}^{*}(x_t)=\infty$, which contradicts the claim that $x_t \in \bar{X}$).

Then, we will show that $\hat{J}(\cdot, \theta)$ is a Lyapunov function for the system. Suppose that $x_t \in \bar{X}\backslash \{0\}$. Applying the optimal Bellman Equation \cite{bertsekas2019reinforcement} for problem \eqref{infinite0} leads to 
\begin{align}\label{bellman}
	J_{\infty}^{*}\left(A x_{t}+B u_{t}^{*}\right)&=J_{\infty}^{*}\left( x_{t}\right)-x^T_tQx_t-u_{t}^{*T}Ru_{t}^{*}\nonumber\\
	&< J_{\infty}^{*}\left( x_{t}\right)-x^T_tQx_t
\end{align}
Combining \eqref{bellman} with \eqref{error2}, we have 
\begin{align}\label{e49}
		&x_{t}^{T} Q x_{t}+\hat{u}_{t}^{* T} R \hat{u}_{t}^{*}+\hat{J}\left(A x_{t}+B \hat{u}_{t}^{*}, \theta\right) \nonumber\\
		 \leq &x_{t}^{T} Q x_{t}+u_{t}^{* T} R u_{t}^{*}+\hat{J}\left(A x_{t}+B u_{t}^{*}, \theta\right) \nonumber\\
		 \leq& x_{t}^{T} Q x_{t}\!+\!u_{t}^{* T} R u_{t}^{*}\!+\!J_{\infty}^{*}\left(A x_{t}\!+\!B u_{t}^{*}\right)\!+\!\zeta J_{\infty}^{*}\left(A x_{t}\!+\!B u_{t}^{*}\right)\nonumber \\
		 < &J_{\infty}^{*}\left(x_{t}\right)+\zeta J_{\infty}^{*}\left(x_{t}\right)-\zeta x_{t}^{T} Q x_{t} \nonumber\\
		 < &\hat{J}\left(x_{t}, \theta\right) + 2\zeta  J_{\infty}^{*}\left(x_{t}\right)-\zeta x_{t}^{T} Q x_{t}
\end{align}
In \eqref{e49}, the first inequality is true since $\hat{u}_{t}^{*}$ is a minimizer of \eqref{onestep}. The second and last inequalities hold due to \eqref{error2}, and the third line is true thanks to the optimal Bellman equation in \eqref{bellman}. Then, combining \eqref{error2} and \eqref{rho1} yields 
\begin{equation}\label{fangsuo}
	\frac{1+\zeta}{\zeta } > 2\mathop {\sup }\limits_{x \in \bar X\backslash \left\{ 0 \right\}} \frac{{\hat J\left( {x,\theta } \right)}}{(1-\zeta){{x^T}Qx}} > 2\mathop {\sup }\limits_{x \in \bar X\backslash \left\{ 0 \right\}} \frac{{J_\infty ^*\left( x \right)}}{{{x^T}Qx}}
\end{equation}
Together with \eqref{fangsuo}, \eqref{e49} implies
\begin{align}\label{e50}
	&\hat{J}\left(A x_{t}+B \hat{u}_{t}^{*}, \theta\right)-\hat{J}\left( x_{t}, \theta\right)\nonumber\\
	 < &\!-\!(1\!+\!\zeta)x_{t}^{T} Q x_{t}\!-\!\hat{u}_{t}^{* T} R \hat{u}_{t}^{*} \!+\! 2\zeta J_{\infty}^{*}\left(x_{t}\right)\!<\!0\; \forall x_t \!\in\! \bar{X}\backslash \{0\}
\end{align}
Let $x_0 \in \bar{X}\backslash \{0\}$ and $x_1,\;x_2,\dots$ be the trajectory of the closed-loop system $x_{t+1}=Ax_t+B\hat{u}_{t}^{*}, t=0,1,\dots$. It follows from \eqref{e50} that the sequence $\hat{J}\left( x_{0}, \theta\right),\;\hat{J}\left( x_{1}, \theta\right),\dots$ is strictly decreasing. Besides, it is easy to check that $\hat{J}\left( 0, \theta\right) = 0$, $\hat{J}\left(x, \theta\right)>0,\;\forall x \in \bar{X}\backslash \{0\}$, and  $\hat{J}\left( \cdot, \theta\right)$ is continuous at the origin, finite in $\bar{X}$. Then, $\hat{J}\left( \cdot, \theta\right)$ is therefore a Lyapunov Function according to \cite[Theorem 7.2]{borrelli2017predictive}. So, the asymptotic stability of the origin for any $x \in \bar{X}$ follows. \qed

As a Lyapunov function is insufficient to guarantee stability for
a constrained system, i.e., the system must stay in an invariant set \cite{jones2009approximate}. In the case of $X_0 \subset \bar{X}$, where the set $X_0$ may not be invariant for the closed-loop system $x_{t+1}=Ax_t+B\hat{u}_{t}^{*}, t=0,1,\dots$, the corresponding stability conditions are described in the following corollary:
\begin{corollary}\label{t7}
	Let $\hat{u}_{t}^{*}$ be the solution to problem \eqref{onestep} with $\mathcal{C} = \mathbb{R}^n$. For a given $X_0  \subset  \bar{X}$, suppose that Assumptions A1-A2 hold. Define a compact set as $\Omega  \triangleq \{ x \in {\mathbb{R}^n}|\;\hat J\left( {x,\theta } \right) \leq \chi \} $ where $\chi  \triangleq \mathop {\inf }\limits_{x \in \partial {X_0}} \hat J\left( {x,\theta } \right)$. If $\zeta$ in \eqref{error2} satisfies
	\begin{equation}\label{rho2}
		\frac{1-\zeta^2}{\zeta } > 2\mathop {\sup }\limits_{x \in \Omega\backslash \left\{ 0 \right\}} \frac{{\hat J\left( {x,\theta } \right)}}{{{x^T}Qx}} ,
	\end{equation}
then $\hat{u}_{t}^{*}$ is recursively feasible for the initial condition $x_0 \in \Omega $, and the origin of the closed-loop system $x_{t+1}=Ax_t+B\hat{u}_{t}^{*}, t=0,1,\dots$ is asymptotically stable with domain of attraction $\Omega $.
\end{corollary}

\textcolor{blue}{For the proof of Corollary \ref{t7}, from the definition of $\Omega $ and the continuity of $\hat J\left( {\cdot,\theta } \right)$, we know that $\Omega \subseteq X_0 $. Following the proof of Theorem \ref{t6}, we can show that for all $x_t \in \Omega$, $\hat{J}\left(A x_{t}+B \hat{u}_{t}^{*}, \theta\right)-\hat{J}\left( x_{t}, \theta\right) <0$ holds. As a result, the set $\Omega $ is positively invariant w.r.t. the closed-loop system $x_{t+1}=Ax_t+B\hat{u}_{t}^{*}$. The recursive feasibility thus follows from $\Omega \subseteq X_0 \subseteq \mathcal{X} $. Similar to  the proof of Theorem \ref{t6}, it can be shown that $\hat{J}(\cdot, \theta)$ is a Lyapunov function for the closed-loop system and stability of the origin follows.}\qed

According to Theorem \ref{t6} and Corollary \ref{t7}, asymptotic stability is achieved if the condition in \eqref{rho1} or \eqref{rho2} holds. \eqref{rho1} or \eqref{rho2} can be satisfied by making $\zeta$ small enough since the right-hand side of \eqref{rho1} or \eqref{rho2} is upper bounded. To prove this, we find that $x^TQx$ tends to zero if and only if $x$ approaches the origin. Let $\mathcal{B} \subseteq \mathcal{O}^\text{LQR}_{\infty}$ be a sufficiently small neighborhood of the origin. It follows that $\hat{J}(x, \theta)= x^T P^* x$ if $x \in  \mathcal{B}$. As a result, $\mathop {\sup }\limits_{x \in \mathcal{B}\backslash \left\{ 0 \right\}} \frac{\hat{J}(x, \theta)}{{{x^T}Qx}} \leq   \frac{{{\lambda _{\max }}\left( {{P^*}} \right)}}{{{\lambda _{\min }}\left( Q \right)}}  $.  Besides, in view of \eqref{bound}, $\zeta$ is determined mainly by $\bar{e}$, $\bar{e}_\mathrm{grad}$, and $d_{i,j}$. The condition \eqref{rho1} or \eqref{rho2} can thereby be guaranteed in two ways. One is to add more hidden units into the neural network so that $\bar{e}$ and $\bar{e}_\mathrm{grad}$ could be smaller according to the universal approximation theorem  \cite{hornik1989multilayer}. Another possibility is to involve more state-value/gradient data to test \eqref{assumption2} so that $d_{i,j}$ is reduced.  

If the data generation strategy in Section 3.2 is used, sometimes it is not straightforward to satisfy Assumption A2 since the positions of the subsequent states $x_{k}^{(i) *},\;i=1,\dots N_x,\;k=1,\dots,N-1$ cannot be determined explicitly. An alternative approach is to only use initial states $\{x^{(i)}\}_{i=1}^{N_{x}}$ with a much larger $N_x$ if we find that a number of regions of the partition $\mathcal{R}_{i, j}$ are not visited by the training points.


Approximation methods of MPC with stability guarantees have been investigated in, e.g., \cite{jones2009approximate,chen2022large,zeilinger2011real}. The stability results in these three papers are based on a common fundamental theorem, stating that if $J_{\infty}\left(x_{0}, \{\hat{u}_{t}^{*}\}^\infty_{t=0} \right)-J_{\infty}^{*}(x_0) $ is less than $x^T_0Qx_0$, $J_{\infty}^{*}(\cdot)$ is still a Lyapunov function for the system controlled by $\{\hat{u}_{t}^{*}\}^\infty_{t=0}$.  Our method, on the contrary, firstly obtains an estimation $\hat{J}\left(\cdot, \theta\right)$ of $J_{\infty}^{*}(\cdot)$. If $\hat{J}\left(\cdot, \theta\right)$ can approximate  $J_{\infty}^{*}(\cdot)$ well (characterized by conditions \eqref{rho1} and \eqref{rho2}), the approximated control law generated by the DP \eqref{onestep} will be inherently stabilizing. In addition, note that the Lyapunov functions in Theorem \ref{t6} and Corollary \ref{t7} are  $\hat{J}\left(\cdot, \theta\right)$, instead of $J_{\infty}^{*}(\cdot)$.

\begin{remark}
	It is worth mentioning that our approach can be extended to the degenerate cases. Even if $J_{\infty}^{*}(\cdot)$ is not continuously differentiable, we can still use the proposed PWQ neural network to approximate it. In addition, all the results of the proposed method apply to the degenerate cases.
\end{remark}

\begin{remark}
	\textcolor{blue}{We investigate model-based learning and assume complete knowledge of the model. Besides, the robustness of our model-based method to uncertainty can be analyzed in the robust MPC framework.}
\end{remark}

\subsection{Complexity analysis}
We analyze the offline storage requirement as well as the online computational complexity of the proposed control scheme, and compare them with other methods, such as implicit MPC, explicit MPC, and the policy approximation methods of MPC \cite{karg2020efficient,chen2022large}.

Storage space is dominated by the number of regions and control laws (for explicit MPC), or the structure of the neural network (for approximate MPC). The storage of some system's parameters, such as $A$, $B$, $\mathcal{X}$, $\mathcal{U}$, $Q$, and $R$, are neglected for consistency.

Implicit MPC does not need to store any data except for some system's parameters. In comparison, to implement explicit MPC, the straightforward way is to store all polyhedral regions $\mathcal{R}_{j},\;j=1,\dots,N_\mathrm{r}$ and all control laws $F_{j} x+g_{j}$, identify online the region that contains the current state, and then choose the corresponding control law. From the explicit control law given in \eqref{pwaforlqr}, explicit MPC requires the storage of $N_\mathrm{r}$ regions and affine feedback laws. Suppose that each region $\mathcal{R}_{j}$ is defined by $n^{(j)}_\mathrm{c}$ constraints. Then, explicit MPC needs to store $\left( {n + 1} \right)\sum\limits_{j = 1}^{{N_\mathrm{r}}} {n_{\text{c}}^{(j)}}  + {N_\mathrm{r}}\left( {mn + m} \right)$ real numbers. As for the proposed method, it needs to construct a PWQ neural network before running the system. The neural network contains 3 parameters: $W,\;b$, and $r$, so the storage of the proposed neural network requires $nM+2M$ real numbers in total. In addition, as for the policy approximation methods reported in \cite{karg2020efficient,chen2022large}, the total storage demand of the neural networks is $\left( {n + n_0 + 1} \right)M + \left( {L - 1} \right)\left( {M + 1} \right)M$, with $n_0 = m$ for \cite{karg2020efficient} and $n_0 = N(m+2n+n_\mathrm{c}+m_\mathrm{c})$ for \cite{chen2022large}, respectively. Here, $L$ denotes the number of hidden layers, and $n_\mathrm{c}$ and $m_\mathrm{c}$ denote the number of constraints specified by $\mathcal{X}$ and $\mathcal{U}$.

Online computation time will be evaluated in terms of floating point operations (flops) for the computations that should be performed online. Implicit MPC needs to solve the QP \eqref{e5} or \eqref{empc} at each time step. Solving \eqref{empc} for a given $x$ requires ${f_\text{QP}}\left( {Nm,\;N\left( {{m_{\text{c}}} + {n_{\text{c}}}} \right)} \right)$ flops in the worst case (\eqref{empc} has no redundant constraints). Here, ${f_\text{QP}}\left( n_\mathrm{D},\;n_\mathrm{I}\right)$ represents the number of flops needed to solve a QP with $n_\mathrm{D}$ decision variables and $n_\mathrm{I}$ linear inequalities. So in an interior-point method, solving \eqref{empc} requires $O\left( {{N^3}{m^3}} \right)$ flops per iteration. In comparison, the number of flops for explicit MPC is $ 2n \sum\limits_{j = 1}^{{N_\mathrm{r}}} {n_{\text{c}}^{(j)}} $ \cite{borrelli2017predictive}.

In our proposed control scheme, solving the DP problem \eqref{onestep} causes computational complexity. In Algorithms 1 and 2, the number of flops to determine $\bar{\mathcal{A}}\left(u^{(s)}\right)$ and to compute the coefficients $\bar{P}\left(\bar{\mathcal{A}}\left(u^{(s)}\right)\right), \bar{q}^{T}\left(\bar{\mathcal{A}}\left(u^{(s)}\right)\right)$ is bounded by $f_\mathrm{act} =M(2n^2+n+m^2+5m+2mn+2)+2mn+m(m+3)/2$ in total. Then, in each iteration Algorithm 1 has to solve 2 different QPs, which need ${f_{{\text{QP}}}}\left( {m,\;n_\mathrm{c} + l_\mathrm{c}+s-1} \right)$ and ${f_{{\text{QP}}}}\left( {m,\;n_s} \right)$ flops. Here, $n_s$ stands for the number of inequalities that define $\mathcal{U}_s$. Algorithm 2 needs only $f_{\text{QP}}\left( m,\;n_\mathrm{c} + {l_\mathrm{c}} \right)$ flops to find $u^{(s)}$ in each iteration. Besides, some other calculation includes the update of $\mathcal{U}_{s}$ ($2m^2+2m-1$ flops per iteration) and the comparison of $\bar{\mathcal{A}}(u^{(s)})$ and $\bar{\mathcal{A}}(u^{(s-1)})$ ($M$ flops per iteration). Therefore, the total number of flops to implement Algorithm 1 is 
\begin{align}\label{flop1}
	f_\mathrm{algo1} =& s_\mathrm{m}(f_\mathrm{act}\!+\!{f_{{\text{QP}}}}\left( {m,\;M \!+\! {n_\mathrm{c}} + {l_\mathrm{c}}} \right)+2m^2+2m-1)\nonumber\\
	&+\sum\limits_{s = 1}^{{s_{\text{m}}}} {{f_{{\text{QP}}}}\left( {m,\;n_s} \right)} 
\end{align}
and the number of flops to calculate the control input by using Algorithm 2 is 
\begin{equation}\label{flop2}
	f_\mathrm{algo2} = s_\mathrm{m}(f_\mathrm{act}+f_{\text{QP}}\left( m,\;n_\mathrm{c} + {l_\mathrm{c}} \right)+M)
\end{equation}
which can be much less than those in Algorithm 1.

\section{Numerical examples}
Three case studies are presented to assess the feasibility and effectiveness of the proposed control scheme. In addition, some other methods including implicit MPC and the policy approximation method of MPC \cite{karg2020efficient}, are also compared with the proposed method in the case studies. The simulations are conducted in MATLAB 2021a. 

\subsection{Example 1: a system with input constraints}
Consider a 2-dimensional linear system with $A = \left[ {\begin{array}{*{20}{c}}
			1&{0.1} \\ 
			{ - 0.1}&1 
	\end{array}} \right],\;B = \left[ {\begin{array}{*{20}{c}}
			1&{0.05} \\ 
			{0.5}&1 
	\end{array}} \right]$ and input constraints $\mathcal{U} = \{u\in \mathbb{R}^2\;|\; ||u||_\infty \leq 0.5\}$. We are interested in stabilizing the system at the origin and meanwhile minimizing the cost $\sum_{k=0}^{\infty} x_{k}^{T} Q x_{k}+u_{k}^{T} R u_{k}$ with $Q = I_2$ and $R = 0.1I_2$. We choose the region of interest $\mathcal{X}_{0}=\left\{x \in \mathbb{R}^{2} \mid\|x\|_{\infty} \leq 3\right\}$. By applying the algorithm in \cite{chmielewski1996constrained}, it can be verified that the vertices of $\mathcal{X}_{0}$ can be steered to an ellipsoidal subset of $\mathcal{O}^\mathrm{LQR}_\infty$ with the horizon $N =10$. By solving the explicit MPC problem \eqref{empc} with $N=10$, the partition of the state space for the optimal control law on the region $\mathcal{X}_{0}$ can be obtained, and is depicted in Fig. 1(a). If some regions share the same form of the control law and their union is convex, the union of these regions is computed and dyed the same color.

\begin{figure}[H]
	\centering  
	\subfigbottomskip=2pt 
	\subfigcapskip=0pt 
	\subfigure[Regions of optimal explicit MPC.]{
		\includegraphics[width=0.42\linewidth]{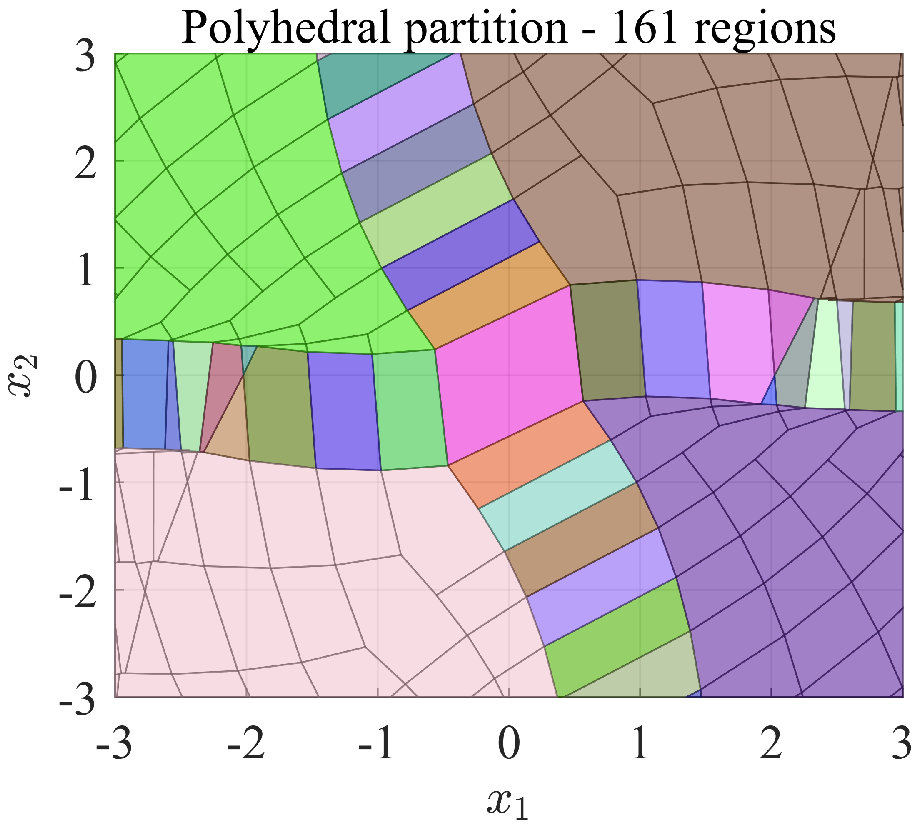}}
	\subfigure[Regions of the PWQ neural network.]{
		\includegraphics[width=0.42\linewidth]{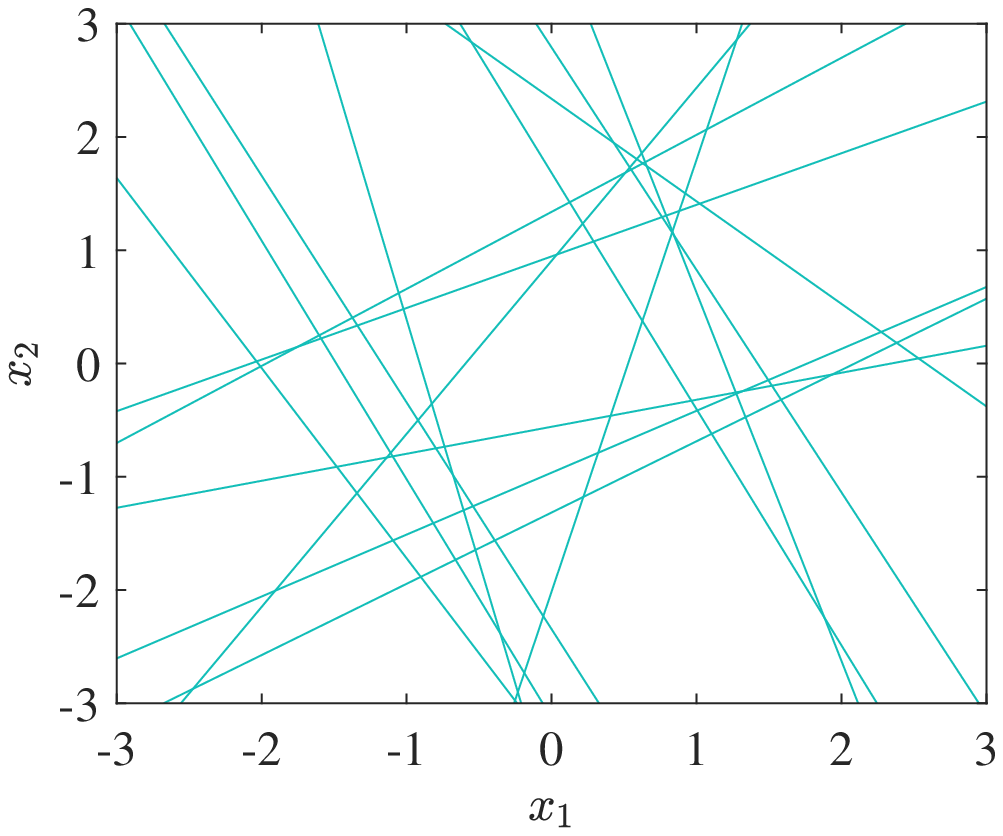}}
	\\
	\subfigure[Comparison of the different types of neural networks.]{
		\includegraphics[width=0.42\linewidth]{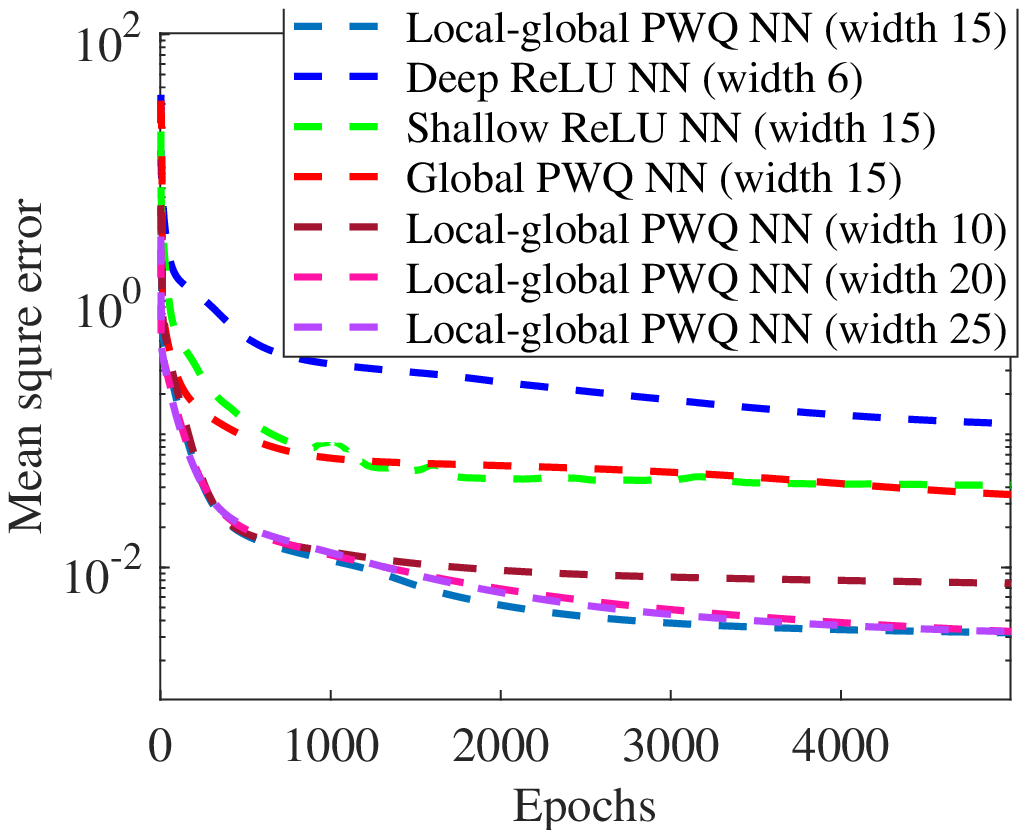}}
	\subfigure[The comparison of the approximated and real value functions.]{
		\includegraphics[width=0.42\linewidth]{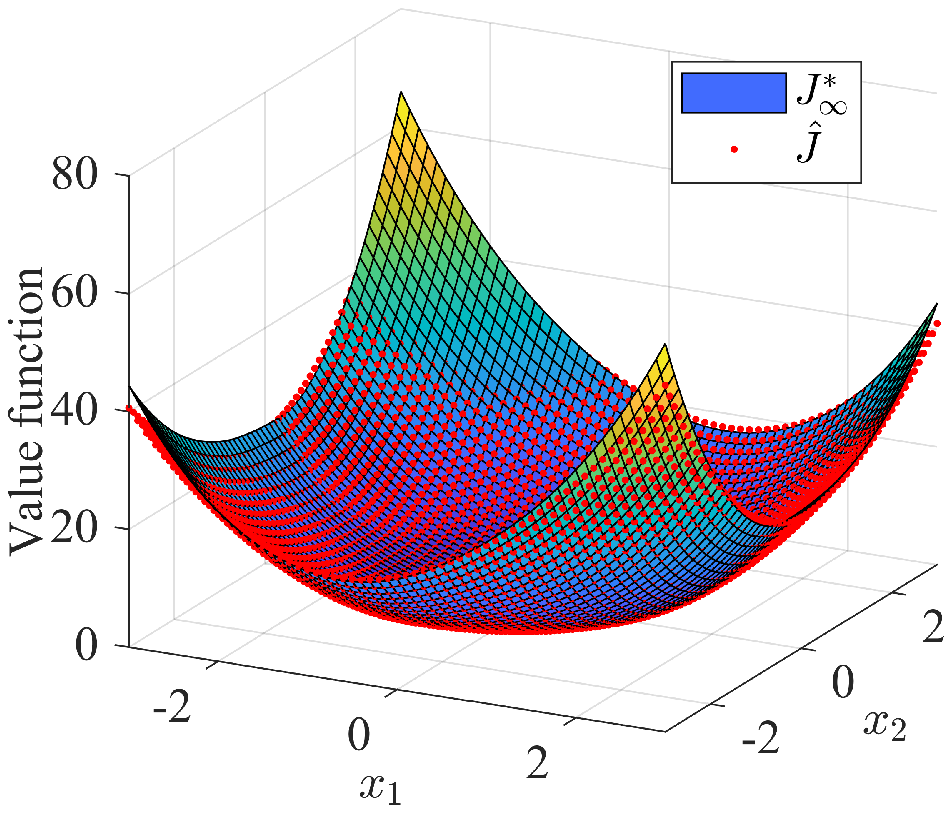}}
\\
	\subfigure[Relative approximation error at training points.]{
		\includegraphics[width=0.42\linewidth]{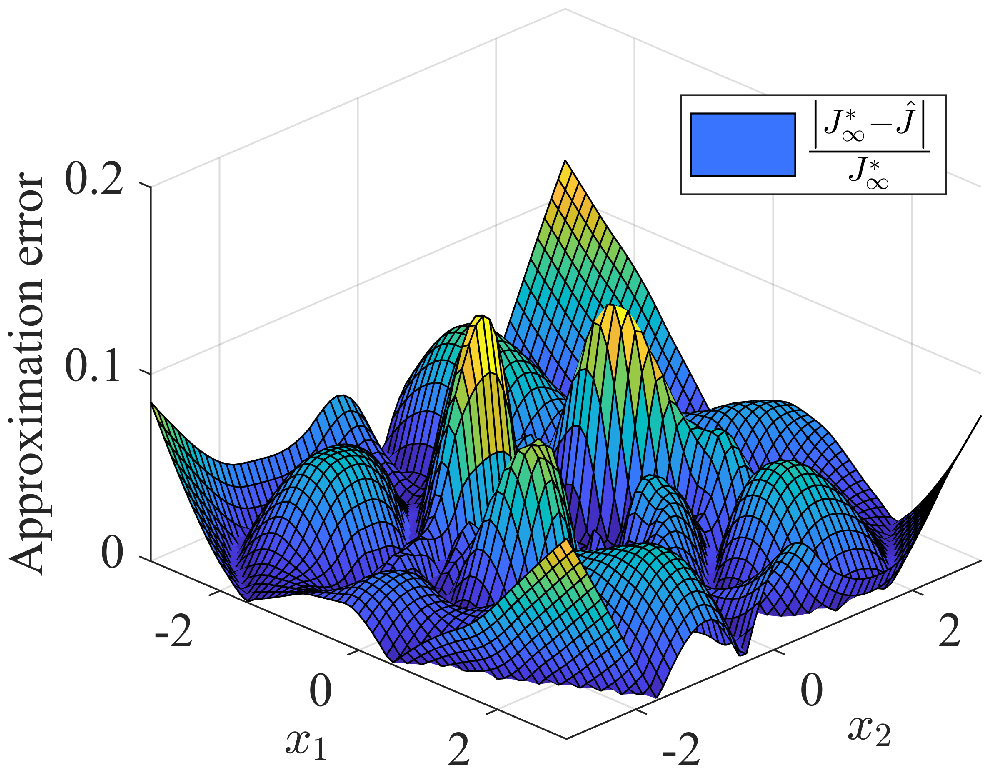}}
	\subfigure[Relative error of the gradient at training points.]{
		\includegraphics[width=0.42\linewidth]{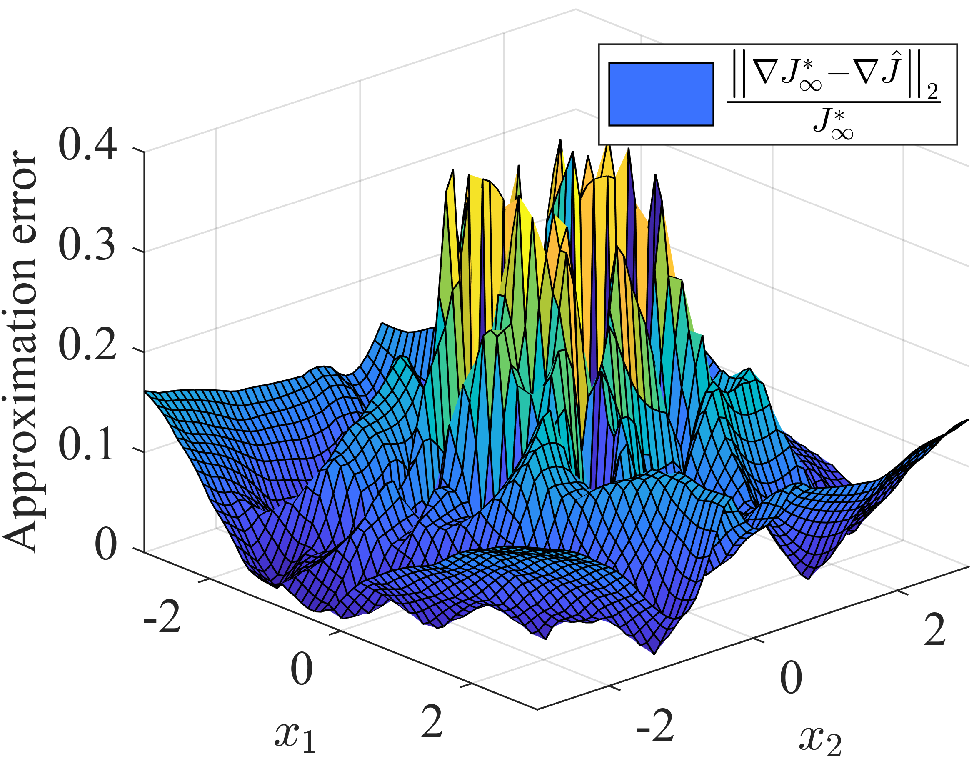}}
	\caption{Performance of the proposed PWQ neural network.}
\end{figure}

To illustrate that the proposed PWQ neural network has a good approximation performance, we compare it with standard ReLU neural networks (with 1 or 2 hidden layers). The widths of the neural networks are taken as 15 (for the proposed and shallow ReLU neural networks) and 6 (for the deep ReLU neural networks), so that all of them have a comparable number of parameters. To prevent the network from falling into local minimums that are not globally optimal, multi-start local optimization \cite{rinnooy1987stochastic} is employed. Particularly, the network is initialized with multiple values of $\theta$ that are randomly selected in the feasible set. We then evaluate the values of the objective function in \eqref{training} and select the initial network parameter that yields the lowest function value. \textcolor{blue}{Besides, a strictly global architecture without the quadratic term $x^T P^* x$ is also compared.} Choosing $\alpha = 0.1$, we compare the absolute mean square errors, i.e., the values of the objective function in \eqref{training} during the training process and show them in Fig. 1(c). From Fig. 1(c), it is observed that the training objectives for the 3 neural networks decrease consistently over epochs, while the proposed approach has the smallest mean square error during the training process. \textcolor{blue}{Additionally, increasing the width does not necessarily improve the approximation ability of the proposed NN.}

The polyhedral partition of the proposed neural network is also depicted in Fig. 1(b). From Fig. 1(b) it is noticed that the partition of the neural network concentrates in the areas where the explicit MPC law changes rapidly. We also compare the output of the PWQ neural network with the real optimal value function in Figs. 1(d)-1(f), from which one can observe that the proposed method is able to closely approximate the optimal value function with a very simple network architecture.

To verify the stability of the closed-loop system, we note that the maximal stabilizable set $\bar{X}$ is open and thereby not computable. So, we concentrate on checking \eqref{rho2}. From Figs. 1(e)-1(f) we can get the bounds $\bar{e} =0.122$ and $\bar{e}_\mathrm{grad} =0.395$. Then, based on \eqref{bound}, we compute the value of the right-hand side of \eqref{bound} at all training points and choose the largest one: $\zeta  =0.188$. Next, the right-hand side of \eqref{rho2} can be computed by solving the following constrained nonlinear problem:
\begin{equation}\label{nonlinearproblem}
	\begin{gathered}
		\max \;\frac{{2\hat J\left( {x,\theta } \right)}}{{{x^T}Qx}} \hfill \\
		{\text{s}}{\text{.t}}{\text{.}}\;\;\epsilon  \leq \hat J\left( {x,\theta } \right) \leq \chi  \hfill \\ 
	\end{gathered} 
\end{equation}
where $\epsilon$ is a small positive constant to avoid singularity. \eqref{nonlinearproblem} returns a maximum 4.904. As a result, it can be readily verified that \eqref{rho2} holds.

In the closed-loop simulation, the proposed method, implicit MPC, and the policy-approximation method of \cite{karg2020efficient} are compared. To implement the policy-approximation method, we construct a deep ReLU neural network with 2 hidden layers, to learn the explicit MPC law. Normally, the width of the deep ReLU neural network should be set as 6 so that it has a comparable number of parameters. However, it shows a bad approximation performance. Therefore, we increase the width of the deep ReLU neural network to 8. \textcolor{blue}{The number of samples for training the policy network is the same as that for training the value network.} When the system is running online, the output of the deep ReLU neural network may violate the input constraint, and thus is projected onto $\mathcal{U}$. All QPs during the simulation are solved via the MATLAB function “quadprog”, and we select an interior-point algorithm when using “quadprog”. We choose the initial point $x_0 = [0\;\;2]^T$ and run the simulation program for 100 steps. The closed-loop behaviors in the first 30 time steps are plotted in Fig. 2. From Fig. 2, it can be seen that the proposed method can properly approximate the MPC controller, and that the corresponding trajectory is stabilized at the origin. In comparison, the policy-approximation method experiences some fluctuations when the state is close to the origin. This is probably because the value of the optimal control input is pretty small if $x$ is around the origin, so a slight approximation error of the policy may lead to drastic changes in the dynamic response. Besides, \textcolor{blue}{we can verify the stability by illustrating the invariance of the sub-level set $\Omega  \triangleq \{ x \in {\mathbb{R}^n}|\;\hat J\left( {x,\theta } \right) \leq \chi \} $. We select some initial states that are at the boundary of $\Omega$, and plot the behavior of the closed-loop system with the proposed controller in Fig. 2(c). It is seen that the trajectories starting from the boundary of $\Omega$ will always stay in $\Omega$, which is computed by Corollary \ref{t7}.}
\begin{figure}\label{xu}
	\centering
	\subfigure[States of the closed loop systems]{\includegraphics[width=110pt, height=100pt,clip]{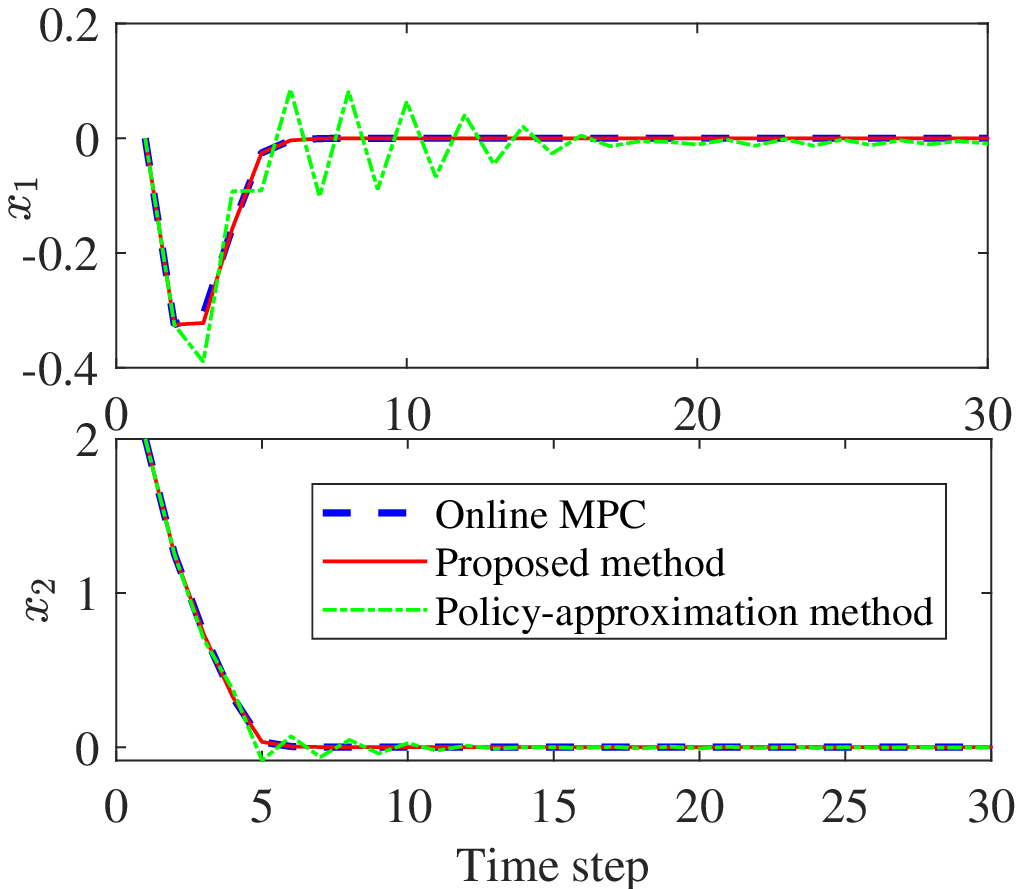}}
	\hfil
	\subfigure[Control inputs]{\includegraphics[width=110pt, height=100pt,clip]{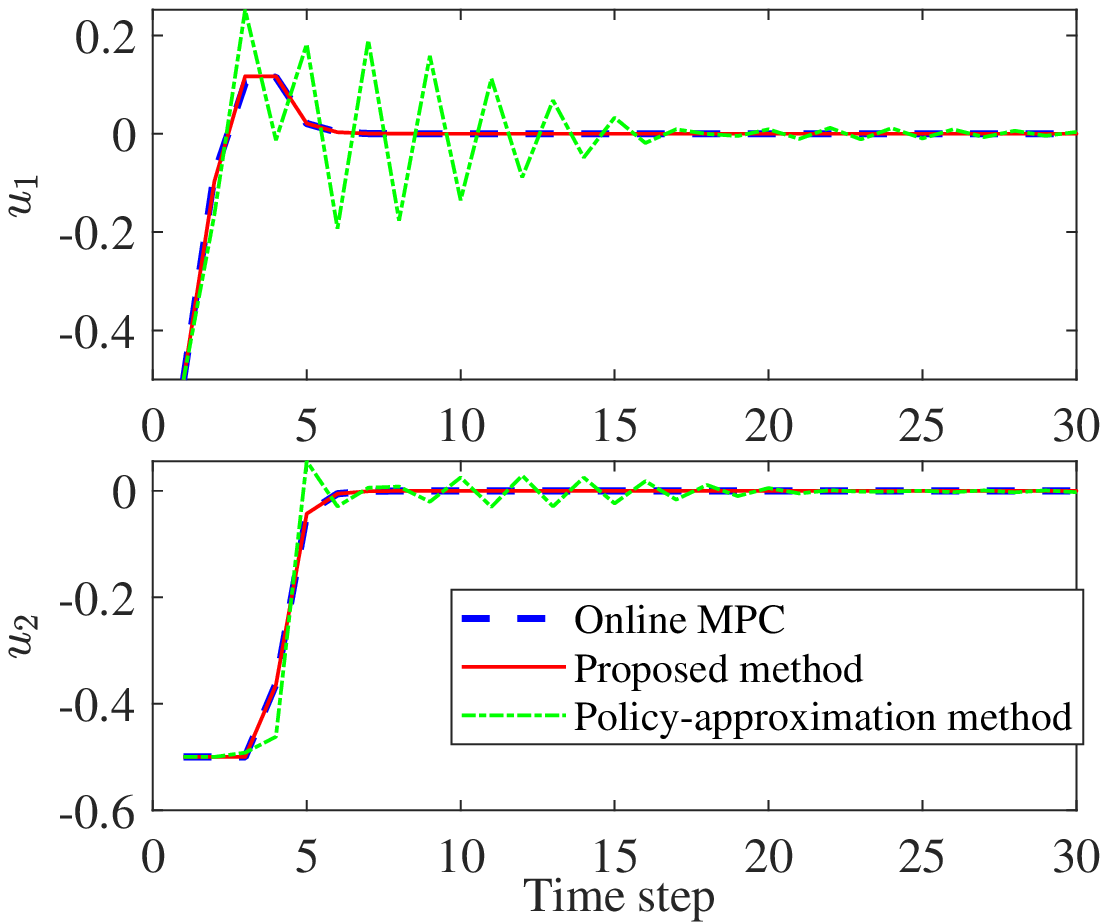}}
	\hfil
	\subfigure[\textcolor{blue}{Verification of the stability.}]{\includegraphics[width=150pt, height=140pt,clip]{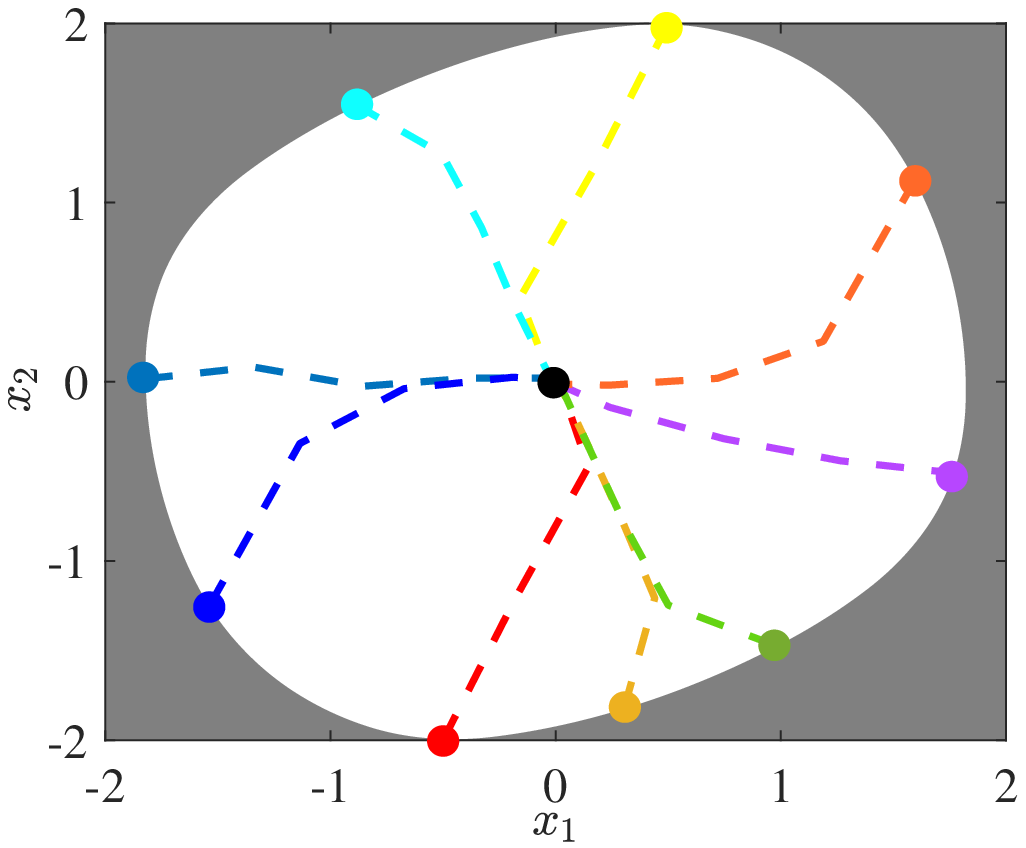}}
	\caption{Closed-loop simulation.}
\end{figure}

\subsection{Example 2: an 8-D system}
\textcolor{blue}{To illustrate the computational performance of the proposed method on larger-size examples we consider the problem of regulating a system of oscillating masses \cite{wang2009fast}. The 8-D system consists of a string of 4 masses that are interconnected by springs with the front and the last mass connected to a wall. The system contains 8 state variables and 2 input variables. The masses have a value of 1 kg. The spring constant is 1 N/m. The constraints are $\|u\|_{\infty} \leq 5$ N, $\|x\|_{\infty} \leq 10$ m. The weight matrices $Q$ and $R$ are set as identity matrices. The MPC horizon is chosen as 20. We select initial states in $\mathcal{X}$ from a uniform distribution and stop until 100 samples that are feasible for MPC are collected. Table 1 shows the average CPU time of the different methods, as well as their total cost after running the system for 100 time steps.} Besides, we also compare the CPU time of the PCP Algorithm, Algorithm 1, and the nonlinear programming algorithm (using the MATLAB function “fmincon” with the interior-point algorithm). For fariness, all QPs are solved via the MATLAB function “quadprog”, using an interior-point algorithm. \textcolor{blue}{The policy-approximation method needs the least simulation time since it just needs to compute a single projection (solve a QP) at each time step. The proposed method with Algorithm 1 requires shorter computation time than online MPC. Besides, compared with the nonlinear solver and the PCP Algorithm, Algorithm 1 significantly reduces the computational complexity. In other aspects, the control law generated by the proposed method is nearly optimal since its total cost is very close to that of MPC, while the total cost of the policy-approximation method is the largest (about 140\% of that of MPC), due to the fluctuations near the origin.}

\begin{table}%
	\begin{small}
		\centering
		\caption{\textcolor{blue}{Comparison of different methods regarding the average CPU time and the total cost in the 8-D system.}}%
		\begin{tabular}{p{120pt}p{40pt}p{40pt}}
			\toprule
			Methods&CPU time &Total cost \\
			\midrule
			Proposed + PCP Algorithm & \textcolor{blue}{1.2300} & \textcolor{blue}{2004.2}\\
			Proposed + Algorithm 1 & \textcolor{blue}{0.2225} & \textcolor{blue}{2004.2}\\
			Proposed + nonlinear solver & \textcolor{blue}{0.7309} & \textcolor{blue}{2004.2}\\
			Online MPC & \textcolor{blue}{1.7279} & \textcolor{blue}{1857.3}\\
			Policy-approximation method & \textcolor{blue}{0.1845}& \textcolor{blue}{2607.5}\\
			\bottomrule
		\end{tabular}
	\end{small}
\end{table}

\textcolor{blue}{For the comparison of the storage demand, in the 2-D example explicit MPC computes a PWA control law with 161 regions. It needs to store 2850 real numbers. In comparison, the proposed NN with width 15 requires the storage of 60 real numbers, which is far less than that for explicit MPC. In the above 8-D example, the explicit solution for the MPC problem is highly complex and cannot be computed within 48 hours. The computation of explicit MPC stopped prematurely due to the available memory being exceeded, with already more than $2\cdot 10^6$ regions stored, while the proposed NN with width 50 needs the storage of 500 real numbers.}

	
	\subsection{Example 3: a system with state and input constraints}
	In the third case study, we concentrate on the case of $X_0 = \bar{X}$. Hence, we consider a linear system with state and input constraints:
	\begin{equation}\label{key}
x_{t+1}=\left[\begin{array}{ll}
				1 & 1 \\
				0 & 1
			\end{array}\right] x_{t}+\left[\begin{array}{l}
				0.4 \\
				0.6
			\end{array}\right] u_{t}, \;\left\|x_{t}\right\|_{\infty} \leq 3,\left|u_{t}\right| \leq 2
	\end{equation}
	where the weight matrices $Q$ and $R$ are taken as $I_2$ and 0.1, respectively. We intend to compare the proposed method with an existing ADP method for constrained linear systems developed in \cite{chakrabarty2019approximate}. With the help of the Multi-Parametric Toolbox, $\bar{X}$ can be efficiently computed, as depicted in Fig. 3.
	\begin{figure}\label{trajectory}
		\centering
		\includegraphics[width=220pt,height=210pt,clip]{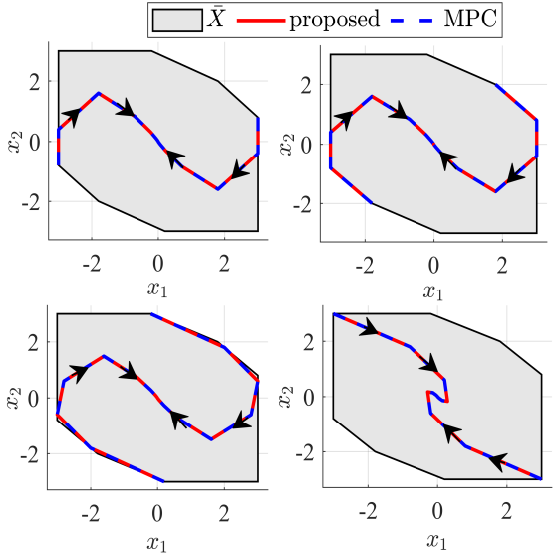}
		\caption{The trajectories of the closed-loop systems controlled by the proposed control law and MPC. The initial states are $ [{ \pm 3\; \pm 0.8} ]^T$ (top left), $ [{ \pm 1.8\; \pm 2} ]^T$ (top right), $[ { \pm 0.2\; \mp 3}  ]^T$ (bottom left), $[{ \pm 3\; \mp 3}  ]^T$ (bottom right).}
	\end{figure}
	
	As in the previous example, it is verified that $N=6$ leads to the equivalence of \eqref{infinite0} and \eqref{empc}. In contrast to using the initial states as well as their subsequent trajectories as training data in the previous example, a 121 $\times$ 121 state data grid is constructed to cover the region $\left\{ {x \in {\mathbb{R}^2}|\;{{\left\| x \right\|}_\infty } \leqslant 3} \right\}$ and only the points contained in $\bar{X}$ are selected as training points. Therefore, the distance between adjacent points is 0.05. These training points are exploited to train the PWQ neural network with width 15. After 15000 training iterations, the value of the objective function in \eqref{training} is less than $10^{-1}$, and we obtain $\bar{e} = 0.0552$, $\bar{e}_\mathrm{grad} =0.355$, and hence $\zeta = 0.0745$. Meanwhile, we can compute the right-hand side of \eqref{rho1}, which equals $9.853$. Consequently, condition \eqref{rho1} is satisfied. The offline analysis hence provides a stability guarantee.
	
	The proposed control law $\hat{u}^*_t$ is expected to be stabilizing for all $x_0$ in $\bar{X}$. To test this, Fig. 3 plots the trajectories of the closed-loop system starting from the vertices of $\bar{X}$. Owing to the convexity of $J^*_\infty(\cdot)$, such vertices have the largest $J^*_\infty(\cdot)$ in their neighborhoods, and therefore the system starting from these vertices is prone to be divergent. However, Fig. 3 shows that the trajectories starting from any vertices converge rapidly to the origin, and the proposed control law significantly approximates the MPC law, which is optimal for the infinite-horizon problem.
	
	Table 2 compares the proposed method (using Algorithm 2) with the ADP method in \cite{chakrabarty2019approximate} and online MPC in terms of total simulation time and total cost in the first 100 time steps. As the ADP method in \cite{chakrabarty2019approximate} is infeasible for all vertices of $\bar{X}$, we choose an initial point $x_0 =[2\;-2]^T$ that is much closer to the origin and run the system in 100 time steps. The proposed method combined with Algorithm 2 takes the least computation time, while the ADP method in \cite{chakrabarty2019approximate} needs the most, which is mainly because \cite{chakrabarty2019approximate} must solve an extra semi-definite program online to estimate the value function. In terms of optimality, the proposed method produces the same solution as MPC since their total costs are equal to each other.
	\begin{table}%
		\begin{small}
			\centering
			\caption{Comparison of different ADP methods and MPC regarding the whole simulation time and the total cost.}%
			\begin{tabular}{p{140pt}p{35pt}p{30pt}}
				\toprule
				Methods&Simulation time (s) &Total cost \\
				\midrule
				Proposed method with Algorithm 2 & 0.142 & 9.956\\
				ADP method in \cite{chakrabarty2019approximate} & 0.270 & 10.367\\
				Online MPC & 0.161 & 9.956\\
				\bottomrule
			\end{tabular}
		\end{small}
	\end{table}
	\section{Conclusions and future work}
	We have developed an ADP control framework for infinite-horizon optimal control of linear systems subject to state and input constraints. Compared to some common neural networks such as ReLU neural networks, the proposed neural network maintains the PWQ property and convexity of the real value function and has a much better approximation performance. These properties and superiority contribute to the reduction of the online computation as well as the construction of explicit stability criteria. Therefore, advantages of our method include low computational requirements, stability assurance, and excellent approximation of the optimal control law. 
	
	The main limitations of the proposed scheme are that the model considered is linear, and that modeling uncertainty is not addressed is the proposed framework. Besides, the supervised learning method needs much offline computation for the preparation of the training data. \textcolor{blue}{Future work includes using temporal difference learning to train the NN, extending the proposed method to the cases where the model is unknown, and considering constrained PWA systems.} 
\vspace{10pt}

\textbf{Acknowledgements:} This paper is part of a project that has received funding from the European Research Council (ERC) under the European Union’s Horizon 2020 research and innovation programme (Grant agreement No. 101018826 - CLariNet).

\bibliographystyle{IEEEtran}        
\bibliography{autosam}           



\appendix

\end{document}